\documentstyle[12pt]{article}

\catcode`\@=11

\let\DOTSI\relax
\def\RIfM@{\relax\ifmmode}
\def\FN@{\futurelet\next}
\newcount\intno@
\def\iint{\DOTSI\intno@\tw@\FN@\ints@}
\def\iiint{\DOTSI\intno@\thr@@\FN@\ints@}
\def\iiiint{\DOTSI\intno@4 \FN@\ints@}
\def\idotsint{\DOTSI\intno@\z@\FN@\ints@}
\def\ints@{\findlimits@\ints@@}
\newif\iflimtoken@
\newif\iflimits@
\def\findlimits@{\limtoken@true\ifx\next\limits\limits@true
 \else\ifx\next\nolimits\limits@false\else
 \limtoken@false\ifx\ilimits@\nolimits\limits@false\else
 \ifinner\limits@false\else\limits@true\fi\fi\fi\fi}
\def\multint@{\int\ifnum\intno@=\z@\intdots@                                
 \else\intkern@\fi                                                          
 \ifnum\intno@>\tw@\int\intkern@\fi                                         
 \ifnum\intno@>\thr@@\int\intkern@\fi                                       
 \int}                                                                      
\def\multintlimits@{\intop\ifnum\intno@=\z@\intdots@\else\intkern@\fi
 \ifnum\intno@>\tw@\intop\intkern@\fi
 \ifnum\intno@>\thr@@\intop\intkern@\fi\intop}
\def\intic@{\mathchoice{\hskip.5em}{\hskip.4em}{\hskip.4em}{\hskip.4em}}
\def\negintic@{\mathchoice
 {\hskip-.5em}{\hskip-.4em}{\hskip-.4em}{\hskip-.4em}}
\def\ints@@{\iflimtoken@                                                    
 \def\ints@@@{\iflimits@\negintic@\mathop{\intic@\multintlimits@}\limits    
  \else\multint@\nolimits\fi                                                
  \eat@}                                                                    
 \else                                                                      
 \def\ints@@@{\iflimits@\negintic@
  \mathop{\intic@\multintlimits@}\limits\else
  \multint@\nolimits\fi}\fi\ints@@@}
\def\intkern@{\mathchoice{\!\!\!}{\!\!}{\!\!}{\!\!}}
\def\plaincdots@{\mathinner{\cdotp\cdotp\cdotp}}
\def\intdots@{\mathchoice{\plaincdots@}
 {{\cdotp}\mkern1.5mu{\cdotp}\mkern1.5mu{\cdotp}}
 {{\cdotp}\mkern1mu{\cdotp}\mkern1mu{\cdotp}}
 {{\cdotp}\mkern1mu{\cdotp}\mkern1mu{\cdotp}}}

\newif\iffirstchoice@
\firstchoice@true
\def\textfonti{\the\textfont\@ne}
\def\textfontii{\the\textfont\tw@}
\def\text{\RIfM@\expandafter\text@\else\expandafter\text@@\fi}
\def\text@@#1{\leavevmode\hbox{#1}}
\def\text@#1{\mathchoice
 {\hbox{\everymath{\displaystyle}\def\textfonti{\the\textfont\@ne}%
  \def\textfontii{\the\textfont\tw@}\textdef@@ T#1}}
 {\hbox{\firstchoice@false
  \everymath{\textstyle}\def\textfonti{\the\textfont\@ne}%
  \def\textfontii{\the\textfont\tw@}\textdef@@ T#1}}
 {\hbox{\firstchoice@false
  \everymath{\scriptstyle}\def\textfonti{\the\scriptfont\@ne}%
  \def\textfontii{\the\scriptfont\tw@}\textdef@@ S\rm#1}}
 {\hbox{\firstchoice@false
  \everymath{\scriptscriptstyle}\def\textfonti
  {\the\scriptscriptfont\@ne}%
  \def\textfontii{\the\scriptscriptfont\tw@}\textdef@@ s\rm#1}}}
\def\textdef@@#1{\textdef@#1\rm\textdef@#1\bf\textdef@#1\sl\textdef@#1\it}
\def\DN@{\def\next@}
\def\eat@#1{}
\def\textdef@#1#2{%
 \DN@{\csname\expandafter\eat@\string#2fam\endcsname}%
 \if S#1\edef#2{\the\scriptfont\next@\relax}%
 \else\if s#1\edef#2{\the\scriptscriptfont\next@\relax}%
 \else\edef#2{\the\textfont\next@\relax}\fi\fi}

\def\Let@{\relax\iffalse{\fi\let\\=\cr\iffalse}\fi}
\def\vspace@{\def\vspace##1{\crcr\noalign{\vskip##1\relax}}}
\def\multilimits@{\bgroup\vspace@\Let@
 \baselineskip\fontdimen10 \scriptfont\tw@
 \advance\baselineskip\fontdimen12 \scriptfont\tw@
 \lineskip\thr@@\fontdimen8 \scriptfont\thr@@
 \lineskiplimit\lineskip
 \vbox\bgroup\ialign\bgroup\hfil$\m@th\scriptstyle{##}$\hfil\crcr}
\def\Sb{_\multilimits@}
\def\endSb{\crcr\egroup\egroup\egroup}
\def\Sp{^\multilimits@}

\newdimen\ex@
\def\rightarrowfill@#1{$#1\m@th\mathord-\mkern-6mu\cleaders
 \hbox{$#1\mkern-2mu\mathord-\mkern-2mu$}\hfill
 \mkern-6mu\mathord\rightarrow$}
\def\leftarrowfill@#1{$#1\m@th\mathord\leftarrow\mkern-6mu\cleaders
 \hbox{$#1\mkern-2mu\mathord-\mkern-2mu$}\hfill\mkern-6mu\mathord-$}
\def\leftrightarrowfill@#1{$#1\m@th\mathord\leftarrow\mkern-6mu\cleaders
 \hbox{$#1\mkern-2mu\mathord-\mkern-2mu$}\hfill
 \mkern-6mu\mathord\rightarrow$}
\def\overrightarrow{\mathpalette\overrightarrow@}
\def\overrightarrow@#1#2{\vbox{\ialign{##\crcr\rightarrowfill@#1\crcr
 \noalign{\kern-\ex@\nointerlineskip}$\m@th\hfil#1#2\hfil$\crcr}}}

\def\overleftarrow{\mathpalette\overleftarrow@}
\def\overleftarrow@#1#2{\vbox{\ialign{##\crcr\leftarrowfill@#1\crcr
 \noalign{\kern-\ex@\nointerlineskip}$\m@th\hfil#1#2\hfil$\crcr}}}
\def\overleftrightarrow{\mathpalette\overleftrightarrow@}
\def\overleftrightarrow@#1#2{\vbox{\ialign{##\crcr\leftrightarrowfill@#1\crcr
 \noalign{\kern-\ex@\nointerlineskip}$\m@th\hfil#1#2\hfil$\crcr}}}
\def\underrightarrow{\mathpalette\underrightarrow@}
\def\underrightarrow@#1#2{\vtop{\ialign{##\crcr$\m@th\hfil#1#2\hfil$\crcr
 \noalign{\nointerlineskip}\rightarrowfill@#1\crcr}}}

\def\underleftarrow{\mathpalette\underleftarrow@}
\def\underleftarrow@#1#2{\vtop{\ialign{##\crcr$\m@th\hfil#1#2\hfil$\crcr
 \noalign{\nointerlineskip}\leftarrowfill@#1\crcr}}}
\def\underleftrightarrow{\mathpalette\underleftrightarrow@}
\def\underleftrightarrow@#1#2{\vtop{\ialign{##\crcr$\m@th\hfil#1#2\hfil$\crcr
 \noalign{\nointerlineskip}\leftrightarrowfill@#1\crcr}}}

\catcode`\@=\active

\def\frac#1#2{{#1 \over #2}}

\def\stackunder#1#2{\mathrel{\mathop{#2}\limits_{#1}}}

\newcount\GRAPHICSTYPE
\def\GRAPHICSPS#1{%
\ifnum\GRAPHICSTYPE=1 language "PS", include "#1"\else%
ps: #1\fi}

\def\graffile#1#2#3#4{\leavevmode\raise -#4 \hbox{%
\raise #3 \hbox{\rule{0.003in}{0.003in}\special{#1}}}%
{\raise -#4 \hbox to #2 {\vrule height#3 width0in depth0in\hfil}}%
}

\def\draftbox#1#2#3#4{\leavevmode\raise -#4 \hbox{\frame{\rlap{\protect\tiny
#1}%
\hbox to #2{\vrule height#3 width0in depth0in\hfil}}}}

\newcount\draft
\def\GRAPHIC#1#2#3#4#5{\ifnum\draft=1 \draftbox{#2}{#3}{#4}{#5}\else%
\graffile{#1}{#3}{#4}{#5}\fi}

\def\addtoLaTeXparams#1{\edef\LaTeXparams{\LaTeXparams #1}}

\def\doFRAMEparams#1{\readFRAMEparams#1\end}
\def\readFRAMEparams#1{%
\ifx#1\end%
\let\next=\relax%
\else%
\ifx#1i%
\dispkind=0%
\fi%
\ifx#1d%
\dispkind=1%
\fi%
\ifx#1f%
\dispkind=2%
\fi%
\ifx#1t%
\addtoLaTeXparams{t}%
\fi%
\ifx#1b%
\addtoLaTeXparams{b}%
\fi%
\ifx#1p%
\addtoLaTeXparams{p}%
\fi%
\ifx#1h%
\addtoLaTeXparams{h}%
\fi%
\let\next=\readFRAMEparams%
\fi%
\next%
}

\def\IFRAME#1#2#3#4#5{\GRAPHIC{#5}{#4}{#1}{#2}{#3}}

\def\DFRAME#1#2#3#4{
  \begin{center}
    \GRAPHIC{#4}{#3}{#1}{#2}{0in}
  \end{center}
}

\def\FFRAME#1#2#3#4#5#6#7{
  \begin{figure}[#1]
    \begin{center}
      \GRAPHIC{#7}{#6}{#2}{#3}{0in}
    \end{center}
    \caption{\label{#5}#4}
  \end{figure}
}

\def\FRAME#1#2#3#4#5#6#7#8{%
\newcount\dispkind%
\def\LaTeXparams{}%
\dispkind=0%
\def\LaTeXparams{}%
\doFRAMEparams{#1}%
\ifnum\dispkind=0%
\IFRAME{#2}{#3}{#4}{#7}{#8}%
\else
  \ifnum\dispkind=1
    \DFRAME{#2}{#3}{#7}{#8}
  \else
    \ifnum\dispkind=2
      \FFRAME{\LaTeXparams}{#2}{#3}{#5}{#6}{#7}{#8}
    \fi
  \fi
\fi
}

\catcode`\@=11

\long\def\QQQ#1#2{}
\def\QTP#1{}
\long\def\QQA#1#2{}

\def\EXPAND#1[#2]#3{}
\def\NOEXPAND#1[#2]#3{}

\def\LaTeXparent#1{}

\@ifundefined{INDEX}{\def\INDEX#1#2{}{}}{}
\@ifundefined{SUBINDEX}{\def\SUBINDEX#1#2#3{}{}{}}{}
\def\initial#1{\bigbreak{\raggedright\large\bf #1}\kern 2pt\penalty3000}

\@ifundefined{abstract}{%
\def\abstract{\if@twocolumn
\section*{Abstract (Not appropriate in this style!)}
\else \small
\begin{center}
{\bf Abstract\vspace{-.5em}\vspace{0pt}}
\end{center}
\quotation
\fi}}{}
\@ifundefined{endabstract}{%
\def\endabstract{\if@twocolumn\else\endquotation\fi}}{}
\@ifundefined{maketitle}{\def\maketitle#1{}}{}
\@ifundefined{affiliation}{\def\affiliation#1{}}{}
\@ifundefined{proof}{}{}
\@ifundefined{newfield}{\def\newfield#1#2{}}{}
\@ifundefined{chapter}{\def\chapter#1{\par(Chapter head:)#1\par }}{}
\@ifundefined{part}{\def\part#1{\par(Part head:)#1\par }}{}
\@ifundefined{section}{\def\part#1{\par(Section head:)#1\par }}{}
\@ifundefined{subsection}{\def\part#1{\par(Subsection head:)#1\par }}{}
\@ifundefined{subsubsection}{\def\part#1{\par(Subsubsection head:)#1\par }}{}
\@ifundefined{paragraph}{\def\part#1{\par(Subsubsubsection head:)#1\par }}{}
\@ifundefined{subparagraph}{\def\part#1{\par(Subsubsubsubsection head:)#1\par
}}{}

\newdimen\theight
\def \Column{%
             \vadjust{\setbox0=\hbox{\scriptsize\quad\quad tcol}%
             \theight=\ht0
             \advance\theight by \dp0    \advance\theight by \lineskip
             \kern -\theight \vbox to \theight{\rightline{\rlap{\box0}}%
             \vss}%
             }}%

\def\qed{\ifhmode\unskip\nobreak\fi\ifmmode\ifinner\else\hskip5\p@\fi\fi
 \hbox{\hskip5\p@\vrule width4\p@ height6\p@ depth1.5\p@\hskip\p@}}
\catcode`@=12 

\makeatletter

\newtheorem{theorem}{Theorem}[section]

\newtheorem{corollary}{Corollary}[section]
\newtheorem{example}{Example}[section]

\author{Michael Semenov-Tian-Shansky$^{1,2}$
Universit\'e de Bourgogne, Dijon,\\ and Steklov Mathematical Institute,
 St.Petersburg
\and
Alexey Sevostyanov\\ 
Steklov Mathematical Institute, St.Petersburg
}

\title{
Classical and Quantum Nonultralocal Systems on the Lattice
}
\date{
}
\QQQ{Language}{
American English
}

\begin{document}

\maketitle
\begin{abstract}
We classify nonultralocal Poisson brackets for 1-dimensional lattice systems
and describe the corresponding regularizations of the Poisson bracket
relations for the monodromy matrix . A nonultralocal quantum algebras on the
lattices for these systems are constructed.For some class of such algebras
an ultralocalization procedure is proposed.The technique of the modified
Bethe-Anzatz for these algebras is developed.This technique is applied to
the nonlinear sigma model problem.
\end{abstract}
\vspace{5cm}
\noindent
$^{1}$Permanent address: Physique Mathematique, Universit\'e de Bourgogne,
21004 Dijon France.\\
e-mail:semenov@pdmi.ras.ru semenov@satie.u-bourgogne.fr\\
$^{2}$Partially supported by the International Science Foundation
\newpage
\section*{ Introduction}

This article is devoted to an old problem, which arose in the beginning of
the development of the Classical Inverse Scattering Method (CISM) \cite{ft}.
An important point of CISM is the calculation of the Poisson brackets
relations for the monodromy matrix of an auxiliary linear problem. This
calculation is usually performed under the technical assumption of
'ultralocality' of the Poisson brackets for local variables (this condition
means simply that the Poisson operator defining the bracket is a
multiplication operator and does not contain any derivations). In many
interesting models this condition is violated, and in this case getting
consistent Poisson brackets relations for the monodromy becomes nontrivial.
Technically, the trouble is that the Frechet derivative of the monodromy has
a discontinuity, and so one has to extend a differential operator to
functions with a jump. It is easy to observe that Poisson operators are
nonultralocal precisely for the models with non-skew-symmetric r-matrices. A
naive calculation of the Poisson brackets for the monodromy in this case
gives:

\begin{equation}
\label{0.1}
\begin{array}{c}
\left\{ M_1,M_2\right\} =aM_1M_2-M_1M_2a, \\
a=\frac 12\left( r-r^{*}\right) .
\end{array}
\end{equation}

This bracket does not satisfy the Jacobi identity, since the skew part of $r$
usually does not satisfy the Yang-Baxter identity (in fact, the bracket( \ref
{0.1}) is inconsistent even if it does). A natural way to regularize the
monodromy brackets in this case has been proposed in \cite{s2}. This method
allows to regularize some (though not all) of the Poisson brackets of the
type (\ref{0.1}). The idea is that to extend the Poisson operator to
functions with a jump one has to add to it a boundary form sensitive to the
jump, which is well in the spirit of the operator extensions theory. In this
article we classify all regularized r-matrices and all regularizations of
this kind using the Belavin-Drinfeld classification theorem for the modified
Yang-Baxter equation \cite{bd}. Unfortunately, our classification is given
in an implicit form because the Belavin-Drinfeld classification theorem
describes solutions of the modified Yang-Baxter equation only up to
automorphisms of the corresponding affine Lie algebra. This fact doesn't
enable to write all regularizations in the explicit form. But we give a
natural way to find  all regularizations. We define the corresponding
quantum algebras by means of the Faddeev-Reshetikhin-Takhtajan approach \cite
{frt}. The same class of Poisson structures and of the corresponding quantum
algebras has been recently studied in a slightly different way by
J.M.Maillet and L.Freidel \cite{maillet} and by S.Parmentier \cite
{parmentier}; to describe them we use the unified approach based on the
notion of the twisted double (cf.\cite{RIMS}, \cite{s3})

The second goal of the present work is to construct quantum nonultralocal
systems on the lattice ,which possess infinite series of conservation laws
and to calculate the spectrum of the corresponding commuting operators. For
this calculation we develop a generalization of the Bethe-Ansatz
construction.

Some words about the contents of this paper.

In section 1 we review the construction of Poisson algebras on the lattice
arising in the study of Lax equations on the lattice with non-ultralocal
Poisson brackets.

In section 2 we remind the main construction of \cite{s2}. We reformulate
the Belavin-Drinfeld classification theorem \cite{bd} in terms of the affine
root systems. This reformulation is convenient for our purposes. We reduce
the classification of regularizations to the search of some class of
solutions of the Yang-Baxter equation on the square of a finite-dimensional
Lie algebra.

In section 3 we discuss the main examples of regularizations and the
corresponding nonultralocal algebras and investigate their algebraic
properties. In particular, we determine their centers; under some additional
conditions it is possible to find a new system of generators of these
algebras which already satisfy {\em local} commutation relations This
ultralocalization procedure has been discussed earlier in \cite{fv2}. We
present new examples of ultralocalization; the new system of generators is
related to the original one by an appropriate quantum lattice gauge
transformation. At the end of section 3 we describe a generalization of the
algebraic Bethe Ansatz for nonultralocal algebras.

In section 4 we apply the technique developed in the previous sections to
the nonlinear sigma model problem. It is well known that integrable models
usually admit several different Poisson structures; the simplest one for the
nonlinear sigma model is associated with its standard Lagrangian
formulation. We were unable to find a regularization of this Poisson
structure; however, the general scheme introduced in section 2 may be
applied to another, and a fairly natural Poisson structure which we
introduce in this section for a nonlinear sigma model with values in an
arbitrary Riemannian symmetric space. We explicitly describe the
corresponding quantum lattice systems . For the n-field (i.e., the sigma
model with values in the unit sphere $S^2)$ we get a representation of the
local quantum lattice Lax operator via the canonical Weyl pairs. It turns
out that the n-field with this Poisson structure is gauge equivalent to the
lattice Sine-Gordon model.

In the conclusion we discuss some open problems.

\section{General construction of lattice algebras}

It is natural to assume that the phase space of a mechanical system
associated with a 1-dimensional lattice $\Gamma ={\bf Z}/N{\bf Z}$ is the
direct product ${\cal M}^N$ of ''1-particle spaces''. In applications to
integrable systems these ''elementary'' phase spaces are parametrized by Lax
matrices and hence are modeled on submanifolds of an appropriate Lie group
(usually, a loop group associated with a finite-dimensional semisimple Lie
group). In simple cases the Poisson structure on ${\cal M}^{N\text{ }}$is
the product structure. (The corresponding Poisson bracket is called {\em %
ultralocal}.) The auxiliary linear problem associated with Lax equations on
the lattice is
\begin{equation}
\label{0.0}\psi _{n+1}=L^n\psi _n.
\end{equation}
The associated monodromy map is the product map
\begin{equation}
\label{001}M:{\em G}^N\rightarrow {\em G}:(L^1,...,L^N)\mapsto
\prod_{n=1}^NL^n.
\end{equation}
It is natural to demand that $M$ is a Poisson map. In ultralocal case this
condition means that $G$ should be a Poisson Lie group. It is interesting
(and also important for applications) to study the most general Poisson
structures on $G^N$ which are compatible with this property of the
monodromy. The corresponding Poisson algebras are referred to as {\em %
lattice algebras}. First examples of nonultralocal lattice algebras appeared
in \cite{RIMS}; further examples and a classification (for finite
dimensional semisimple Lie algebras) appeared in \cite{maillet}, \cite{afs},
\cite{parmentier}. In this section we briefly recall the construction of
lattice algebras using the approach proposed in \cite{RIMS}, \cite{s2}.

Fix an affine Lie algebra ${\bf g}$ with a normalized invariant bilinear
form $\left\langle \cdot ,\cdot \right\rangle $.\ It is well known that $%
{\bf g}$ admits the structure of a quasitriangular Lie bialgebra (the
corresponding classical r-matrices are listed in \cite{bd}). Put ${\bf d}=%
{\bf g}\oplus {\bf g}{\em .}$ We define the bilinear invariant form on the
square of ${\bf g}$ in the following way:

\begin{equation}
\label{1.1}\left\langle \left\langle \left( X_1,Y_1\right) ,\left(
X_2,Y_2\right) \right\rangle \right\rangle =\left\langle
X_1,X_2\right\rangle -\left\langle Y_1,Y_2\right\rangle ,
\end{equation}

so that the diagonal subalgebra is isotropic. As a Lie algebra, ${\bf d}$ is
isomorphic to the double of ${\bf g}.$ (This isomorphism does not depend on
a particular choice of the r-matrix.) Hence ${\bf d}$ carries a natural
r-matrix, the r-matrix of the double; for our present goals, however, we
shall need {\em arbitrary} classical r-matrices on ${\bf d}$ which define on
it the structure of a quasitriangular Lie bialgebra. In other words, we are
interested in r-matrices which are skew with respect to( \ref{1.1}) and
satisfy the modified classical Yang-Baxter equation on ${\bf g}\oplus {\bf g}%
{\em .}$

Let ${\cal \ R}$ be such a solution; it may be written in the block form:
\begin{equation}
\label{1.2}{\cal R=}\ \ \left(
\begin{array}{cc}
A & B \\
B^{*} & D
\end{array}
\right) ,\ A^{*}=-A,\ D^{*}=-D.
\end{equation}

For $\varphi \in Fun\left( {\em D}\right) ,{\em D}={\em G}\times {\em G}$
let ${\rm D}\varphi ,$ ${\rm D}^{\prime }\varphi \in \left( {\bf g}\oplus
{\bf g}\right) ^{*}$ be the left and right derivatives of $\varphi $:

\begin{equation}
\label{1.4}
\begin{array}{c}
\left\langle \left\langle
{\rm D}\varphi \left( g\right) ,X\right\rangle \right\rangle =\frac
d{dt}\mid _{t=0}\varphi \left( e^{tX}g\right) , \\ \left\langle \left\langle
{\rm D}^{\prime }\varphi \left( g\right) ,X\right\rangle \right\rangle
=\frac d{dt}\mid _{t=0}\varphi \left( ge^{tX}\right) , \\ g\in {\em D,\ }%
X\in {\bf g}\oplus {\bf g.}
\end{array}
\end{equation}

It is well known that for any two solutions ${\cal R},{\cal R}^{\prime }$ of
the modified classical Yang-Baxter equation the bracket
\begin{equation}
\label{1.4.1}\left\{ \varphi ,\psi \right\} _{{\cal R}{\em ,}{\cal R}%
^{\prime }}=\left\langle \left\langle {\cal R}_1{\rm D}\varphi ,{\rm D}\psi
\right\rangle \right\rangle +\left\langle \left\langle {\cal R}_2{\rm D}%
^{\prime }\varphi ,{\rm D}^{\prime }\psi \right\rangle \right\rangle
\end{equation}
satisfies the Jacobi identity. Let us take, in particular, ${\cal R}_1={\cal %
R},{\cal R}_2=\pm {\cal R}$ we get the following important brackets

\begin{equation}
\label{1.5}\left\{ \varphi ,\psi \right\} _{{\em D}_{\pm }}=\left\langle
\left\langle {\cal R}{\rm D}\varphi ,{\rm D}\psi \right\rangle \right\rangle
\pm \left\langle \left\langle {\cal R}{\rm D}^{\prime }\varphi ,{\rm D}%
^{\prime }\psi \right\rangle \right\rangle .
\end{equation}

We denote by ${\em D}_{\pm }$ the group ${\em D}$ with the bracket $\left\{
\cdot ,\cdot \right\} _{{\em D}_{\pm }}$ . The bracket $\left\{ ,\right\} _{%
{\em D}_{-}}$ equips ${\em D}$ with the structure of a Poisson-Lie group,
while the ''+'' sign corresponds to an almost nondegenerate Poisson
structure on ${\em D}_{+}$. (It is symplectic on an open cell in ${\em D}$
containing the unit element, see \cite{aleksmalkin} for the description of
the symplectic leaves of ${\em D}_{+}.$)

Multiplication map {\em D}$\times {\em D}\rightarrow {\em D}$ defines a
Poisson group action {\em D}$_{-}\times {\em D}_{+}\rightarrow {\em D}_{+};$
its restriction to the diagonal subgroup {\em G}$\subset {\em D}$ is {\em %
admissible} \cite{RIMS}, and hence it is possible to perform Poisson
reduction over the action of {\em G}$.$ The quotient space is canonically
identified with ${\em G}$ itself; in fact, it is clear that the map $\pi :%
{\em D}\rightarrow {\em G}:\left( g_1,g_2\right) \mapsto g_1g_2^{-1}$ is
constant on the right coset classes of {\em G}$.$

To calculate the explicit form of the quotient Poisson structure on ${\em G}$
choose $\varphi \in Fun\left( {\em G}\right) $ and put $\hat \varphi =\pi
^{*}\varphi ;$ let $\nabla \varphi ,\nabla ^{\prime }\varphi $ be the left
and right derivatives of $\varphi .$

\begin{equation}
\label{1.8}
\begin{array}{c}
\left\langle \nabla \varphi \left( g\right) ,X\right\rangle =\frac d{dt}\mid
_{t=0}\varphi \left( e^{tX}g\right) , \\
\left\langle \nabla ^{\prime }\varphi \left( g\right) ,X\right\rangle =\frac
d{dt}\mid _{t=0}\varphi \left( ge^{tX}\right) , \\
g\in {\em G,}\text{ }X\in {\bf g}{\em .}
\end{array}
\end{equation}

Then
\begin{equation}
\label{1.9}{\rm D}\hat \varphi \left( g_1,g_2\right) =\left( \nabla \varphi
\left( g_1g_2^{-1}\right) ,\nabla ^{\prime }\varphi \left(
g_1g_2^{-1}\right) \right) .
\end{equation}

After a short computation this yields:

\begin{equation}
\label{1.10}\left\{ \varphi ,\psi \right\} _{{\em G}}=\left\langle A\nabla
\varphi ,\nabla \psi \right\rangle -\left\langle D\nabla ^{\prime }\varphi
,\nabla ^{\prime }\psi \right\rangle +\left\langle B\nabla ^{\prime }
\varphi ,\nabla \psi \right\rangle -\left\langle B^{*}\nabla \varphi ,\nabla
^{\prime }\psi \right\rangle .
\end{equation}

In general, this Poisson structure is degenerate.

Suppose that $\tau $ is an automorphism of ${\bf g};$then $\tau \oplus \tau $
is an automorphism of ${\bf g}\oplus {\bf g}.$ Let us assume that $\tau
\oplus \tau $ commutes with ${\cal R}$ . To twist the r-matrix on ${\bf d}$
we shall use another extension of $\tau $ to ${\bf d};$ namely, we put $\hat
\tau =\tau \oplus \tau ^{-1}.$ Put
\begin{equation}
\label{1.11}{\cal R}^\tau =\hat \tau {\cal R}\hat \tau ^{-1}=\left(
\begin{array}{cc}
A & B\tau ^{-1} \\
\tau B^{*} & D
\end{array}
\right) .
\end{equation}

${\cal R}$ $^\tau $ satisfies the Yang-Baxter equation. Put ${\cal R}_1=%
{\cal R}^\tau ,{\cal R}_2={\cal R}$ in (\ref{1.4.1}) and denote by ${\em D}%
_\tau $ the group ${\em D}$ with the corresponding Poisson structure$:$%
\begin{equation}
\label{1.12}\left\{ \varphi ,\psi \right\} _{{\em D}_\tau }=\left\langle
\left\langle {\cal R}^\tau {\rm D}\varphi ,{\rm D}\psi \right\rangle
\right\rangle +\left\langle \left\langle {\cal R}{\rm D}^{\prime }\varphi ,%
{\rm D}^{\prime }\psi \right\rangle \right\rangle .
\end{equation}

(If ${\cal R}$ is the r-matrix of the double of ${\bf g}$, the group {\em D}$%
_\tau $ is usually referred to as the {\em twisted double}.) This Poisson
structure on ${\em D}_\tau $ also admits reduction with respect to the
action of the diagonal subgroup; the quotient structure on {\em G}$=\pi (%
{\em D}_\tau )$ is given by

\begin{equation}
\label{1.14}\left\{ \varphi ,\psi \right\} _\tau =\left\langle A\nabla
\varphi ,\nabla \psi \right\rangle -\left\langle D\nabla ^{\prime }\varphi
,\nabla ^{\prime }\psi \right\rangle +\left\langle B\tau ^{-1}\nabla
^{\prime }\varphi ,\nabla \psi \right\rangle -\left\langle\tau B^{*}\nabla
\varphi ,\nabla ^{\prime }\psi \right\rangle .
\end{equation}

In particular, let us apply this construction to the group ${\sf G}={\em G}%
^N;$in this case ${\sf D}=\left( {\em G}\times {\em G}\right) ^N,$ and $\tau
$ is the cyclic shift in the direct sum $\bigoplus\limits_1^N{\bf g}$. Let
\begin{equation}
\label{rmat}r=\left(
\begin{array}{cc}
A & B \\
B^{*} & D
\end{array}
\right)
\end{equation}

be a solution of the modified classical Yang-Baxter equation on ${\bf g}%
\oplus {\bf g}$; put ${\cal R}=\bigoplus\limits_1^Nr.$ Evidently, ${\cal R}$
commutes with $\tau \oplus \tau .$ To describe the resulting lattice Poisson
algebra it is convenient to introduce tensor notations. Fix an exact matrix
representation $\rho _{V\text{ }}$ of ${\em G}$ and denote

\begin{equation}
\label{1.15}
\begin{array}{c}
L^n=\rho _{V
\text{ }}\left( g_n\right) ,L_1^n=L^n\otimes I,L_2^n=I\otimes L^n, \\
g=\left( g_{1,\ldots ,}g_n\right) \in {\em G}^N.
\end{array}
\end{equation}

The reduced Poisson brackets of the matrix coefficients of $L^n$ have the
form:

\begin{equation}
\label{1.16}
\begin{array}{c}
\left\{ L_1^n,L_2^n\right\} =-AL_1^nL_2^n+L_1^nL_2^nD, \\
\left\{ L_1^n,L_2^{n+1}\right\} =L_1^nB^{*}L_2^{n+1}, \\
\left\{ L_1^n,L_2^m\right\} =0,\left| n-m\right| \geq 2.
\end{array}
\end{equation}

Here we denote $\left( \rho _V\otimes \rho _V\right) A$ as well as $A$ $,$%
etc. The main property of the Poisson bracket (\ref{1.16}) is given by the
following assertion:

\begin{enumerate}
\begin{theorem}
\- \cite{s3}

Equip ${\sf G}={\em G}^N$ with the Poisson structure (\ref{1.10}); then the
monodromy map
$$
M:{\sf G}_\tau \rightarrow {\em G},\ M\left( g_1,\ldots ,g_N\right)
=g_1\cdot \ldots \cdot g_N
$$
is a Poisson mapping if and only if the r-matrix (\ref{rmat}) satisfies the
additional constraint
\begin{equation}
\label{1.3}A+B=B^{*}+D.
\end{equation}
In that case the Poisson structure in the target space of the monodromy map
is given by (\ref{1.10}).
\end{theorem}
\end{enumerate}

In tensor notations we have the following brackets for $M:$

\begin{equation}
\label{1.17}\left\{ M_1,M_2\right\} =-AM_1M_2+M_1M_2D+M_1B^{*}M_2-M_2BM_1
\end{equation}

Later we shall describe symplectic leaves of the bracket (\ref{1.16}) in the
main examples. In the particular case when $B=0,A=D\ $the bracket is
ultralocal. The reader may keep in mind this possibility as a degenerate
case.

The remainder of this section is devoted to the quantization of the Poisson
brackets (\ref{1.16}), (\ref{1.17}). It may be easily performed on the lines
of \cite{frt} provided that we know the quantum R-matrix which corresponds
to the chosen classical r-matrix on ${\bf d}.$ More precisely, let $U_q({\bf %
d;}{\cal R})$ be the quantized universal enveloping algebra of ${\bf d}$
which corresponds to ${\cal R}$ \cite{d1}; note that its description is not
quite obvious since in the existing literature only the standard algebras $%
U_q({\bf d;}{\cal R})$ which correspond to simplest solutions of the
classical Yang-Baxter equation are usually considered. It is widely believed
that all solutions from the Belavin-Drinfeld list \cite{bd} give rise to
quasitriangular Hopf algebras. (Examples in section 3 below give evidence to
support this belief.) Assuming that the algebra $U_q({\bf d;}{\cal R})$
exists, let

\begin{equation}
\label{1.18}{\cal R}_q=\left(
\begin{array}{cc}
A_q & B_q \\
B_q^{*} & D_q
\end{array}
\right) \in U_q({\bf d;}{\cal R})\otimes U_q({\bf d;}{\cal R})
\end{equation}

be its universal quantum R-matrix. We omit the explicit form of the
relations in the algebra \-$U_q({\bf d;}{\em R}).$ Let $\rho _V$ be some
representation of the algebra $U_q({\bf d;}{\cal R})$ in the space $V$ and
let

\begin{equation}
\label{1.19}{\cal R}_q^{VV}=\ \left( \rho _V\otimes \rho _V\right) {\cal R}%
_q=\left(
\begin{array}{cc}
A_q & B_q \\
B_q^{*} & D_q
\end{array}
\right)
\end{equation}

The following theorem is parallel to the description of the twisted quantum
double and of the lattice current algebra \cite{s3}, \cite{afs}.

\begin{theorem}
The free algebra $Fun_q^{{\cal R}}\left( {\sf G}_\tau \right) $ generated by
the matrix elements of the matrices $L^n\in Fun_q^{{\cal R}}\left( {\sf G}%
_\tau \right) \otimes End\left( V\right) $, satisfying the following
relations:
\begin{equation}
\label{1.20}
\begin{array}{c}
A_qL_1^nL_2^n=L_2^nL_1^nD_q \\
L_1^nB_q^{*^{-1}}L_2^{n+1}=L_2^{n+1}L_1^n
\end{array}
,
\end{equation}

is the quantization of the Poisson algebra (\ref{1.16}).
\end{theorem}

\begin{theorem}
The free algebra $Fun_q^{{\cal R}}\left( {\em G}\right) $ generated by the
matrix elements of the matrix $M\in Fun_q^{{\cal R}}\left( {\em G}\right)
\otimes End\left( V\right) $, satisfying the relations:
\begin{equation}
\label{1.21}A_qM_1B_q^{*^{-1}}M_2=M_2B_q^{-1}M_1D_q
\end{equation}
is the quantization of the Poisson algebra (\ref{1.17}).
\end{theorem}

Finally, we formulate the quantum version of theorem 1.1

\begin{theorem}
The map%
$$
\begin{array}{c}
M:Fun_q^{
{\cal R}}\left( {\sf G}_\tau \right) \rightarrow Fun_q^{{\cal R}}\left( {\em %
G}\right) , \\ \left( L^1,\ldots ,L^N\right) \mapsto L^1\cdot \ldots \cdot
L^N
\end{array}
$$

is an homomorphism of algebras.
\end{theorem}

The algebras (\ref{1.20}), (\ref{1.21}) are the principal objects of our
investigation.

\section{Regularization of nonultralocal Poisson brackets}

The goal of this section is to link the construction of lattice algebras
with Hamiltonian systems on coadjoint orbits of current algebras. This
approach is outlined in \cite{s2} where a regularization procedure for the
Poisson brackets of the monodromy matrix is proposed which matches naturally
with lattice Poisson algebras described in section 1. This will also allow
us to construct consistent lattice approximations of nonultralocal systems
on the circle.

We remind some details of the construction of dynamical systems on coadjoint
orbits \cite{s1}, \cite{ft}.

Let ${\bf G}=C^\infty \left( S^1,{\bf g}\right) $ be a current algebra with
the values in some affine Lie algebra ${\bf g}${\em \ }. Let us define the
invariant scalar product on ${\bf G}${\bf :}

\begin{equation}
\label{2.1}\left( X,Y\right) =\int\limits_0^{2\pi }\left\langle
X,Y\right\rangle dz,
\end{equation}

where $X,Y\in {\em \ }{\bf G},\left\langle \cdot ,\cdot \right\rangle $ is a
invariant bilinear form on ${\bf g}${\em . } Let $\widehat{{\bf G}}$ be the
central extension of the algebra ${\bf G}$ which corresponds to the 2-cocycle

\begin{equation}
\label{2.2}\omega \left( X,Y\right) =\left( X,\partial _zY\right) .
\end{equation}

By definition, $\widehat{{\bf G}}$ is the set of pairs $\left( X,a\right)
,X\in {\bf G},a\in {\bf C}$ with the commutator

\begin{equation}
\label{2.3}\left[ \left( X.a\right) ,\left( Y,b\right) \right] =\left(
\left[ X,Y\right] ,\omega \left( X,Y\right) \right) .
\end{equation}

If $r$ is a solution of the modified classical Yang-Baxter equation on ${\bf %
g}$, we put as usual%
$$
\left[ X,Y\right] _r=\left[ rX,Y\right] +\left[ X,rY\right] .
$$
Let ${\bf g}_r$ be the algebra ${\bf g}$ equipped with this bracket. Put $%
{\bf G}_r=C^\infty \left( S^1,{\bf g}_r\right) ;$ it is easy to see that
$$
\omega _r\left( X,Y\right) =\omega \left( rX,Y\right) +\omega \left(
X,rY\right)
$$
is a 2-cocycle on ${\bf G}_r$; thus we may define the second structure of a
Lie algebra on $\widehat{{\bf G}}$

\begin{equation}
\label{2.5}\left[ \left( X,a\right) ,\left( Y,b\right) \right] _r=\left(
\left[ X,Y\right] _r,\omega _r\left( X,Y\right) \right) .
\end{equation}

In this formula it is not assumed that $r$ is skew-symmetric with respect to
the scalar product $\left\langle \cdot ,\cdot \right\rangle $. (In fact, if
it is, the cocycle $\omega _r$ vanishes identically.)

Let $\ \widehat{{\bf G}}^{*}$ be the dual space of $\widehat{{\bf G}}$ ;
using the inner product (\ref{2.1}) we may identify it with ${\bf G}\oplus
{\bf C.}$ The Poisson bracket used in the CISM is the Lie-Poisson bracket
which corresponds to the commutator (\ref{2.5}). The variable $e\in {\bf C}$
is central with respect to this bracket. If $X_\varphi \in {\em \ }\widehat{%
{\bf G}}$ is a derivative of a function $\varphi \in Fun\left( {\em \ }%
\widehat{{\bf G}}^{*}\right) :$

\begin{equation}
\label{2.6}
\begin{array}{c}
\left( \left( X_\varphi ,X\right) \right) =\frac d{dt}\mid _{t=0}\varphi
\left( L+tX\right) , \\
X,L\in
\widehat{{\bf G}}^{*}, \\ \text{here }\left( \left( \cdot ,\cdot \right)
\right) \text{ is a natural pairing between }\widehat{{\bf G}}\text{ and }%
\widehat{{\bf G}}^{*}.
\end{array}
\end{equation}

then

\begin{equation}
\label{2.7}\left\{ \varphi ,\psi \right\} \left( L,e\right) =\left( \left(
\left( L,e\right) ,\left[ X_\varphi ,X_\psi \right] _r\right) \right) .
\end{equation}

Without loss of generality we may assume that $e=1$ and suppress it in the
notations. The bracket (\ref{2.7}) may be represented as the bilinear form
of the Poisson operator:

\begin{equation}
\label{2.8}{\cal H}=adL\circ r+r^{*}\circ adL-\left( r+r^{*}\right) \partial
_z,
\end{equation}

\begin{equation}
\label{2.9}\left\{ \varphi ,\psi \right\} \left( L\right) =\left( {\cal H}%
X_\varphi ,X_\psi \right) .
\end{equation}

The operator ${\cal H}$ is unbounded, so the formula (\ref{2.9}) requires
some caution when the gradients are not smooth on the circle. This is
precisely the case for the Poisson brackets of the monodromy matrix. Let $%
\psi $ be the fundamental solution of the equation:

\begin{equation}
\label{2.10}\partial _z\psi =L\psi
\end{equation}

normalized by $\psi (0)=I;$ then the monodromy matrix is equal to

\begin{equation}
\label{2.11}M=\psi \left( 2\pi \right) \in {\em G.}
\end{equation}

Fix $\Phi \in Fun({\em G})$. According to \cite{ft}, the Frechet derivative
of the functional $L\mapsto \Phi \left( M\left[ L\right] \right) $ is given
by

\begin{equation}
\label{2.12}X_\Phi \left( z\right) =\psi \left( z\right) \nabla ^{\prime
}\Phi \left( M\right) \psi \left( z\right) ^{-1}
\end{equation}

and in general is discontinuous at $z=0:$

\begin{equation}
\label{2.13}
\begin{array}{c}
X_\Phi \left( 0\right) =\nabla ^{\prime }\Phi \left( M\right) , \\
X_\Phi \left( 2\pi \right) =\nabla \Phi \left( M\right) .
\end{array}
\end{equation}

To regularize Poisson brackets of the monodromy we shall use an idea
borrowed from the theory of self-adjoint extensions \cite{s2}. Let $%
\triangle :C^\infty (\left[ 0,2\pi \right] ;{\bf g})\rightarrow {\bf g}%
\oplus {\bf g}$ be the map which associates to a function on $\left[ 0,2\pi
\right] $ its boundary values,

\begin{equation}
\label{2.14}\triangle X_\varphi =\left(
\begin{array}{c}
X_\varphi \left( 0\right) \\
X_\varphi \left( 2\pi \right)
\end{array}
\right) .
\end{equation}

Choose $B\in End\left( \stackrel{\circ }{\bf g}\oplus \stackrel{\circ }{\bf g%
}\right) ,$ here $\stackrel{\circ }{\bf g}\subset {\bf g}$ is the spreading
finite-dimensional Lie algebra which corresponds to an affine Lie algebra $%
{\bf g}$ ; we extend operator $B$ to the space ${\bf g}\oplus {\bf g}$ as a
zero operator outside $\stackrel{\circ }{\bf g}\oplus \stackrel{\circ }{\bf g%
}$ and define the regularized Poisson bracket in the following way:

\begin{equation}
\label{2.15}\left\{ \varphi ,\psi \right\} \left( L,1\right) =\frac 12\left(
\left( {\cal H}X_\varphi ,X_\psi \right) -\left( {\cal H}X_\psi ,X_\varphi
\right) \right) +\left\langle \left\langle B\triangle X_\varphi ,\triangle
X_\psi \right\rangle \right\rangle .
\end{equation}

The bracket (\ref{2.15}) must coincide with the bracket (\ref{2.9}) on
smooth functions, hence $B$ must satisfy the condition:

\begin{equation}
\label{2.16}\left\langle \left\langle B\left(
\begin{array}{c}
X \\
X
\end{array}
\right) ,\left(
\begin{array}{c}
Y \\
Y
\end{array}
\right) \right\rangle \right\rangle =0.
\end{equation}

The additional restriction on $B$ imposed in \cite{s2} follows from the
study of the linearized bracket for the monodromy (\ref{2.15}) for $%
M\rightarrow 1$; it is natural to demand that this linearized bracket should
coincide with the one defined by $r$. This gives, after a short computation:

\begin{equation}
\label{2.18}\left\{ \Phi ,\Psi \right\} \left( M\right) =\left\langle
\left\langle {\cal R}\left(
\begin{array}{c}
\nabla \Phi \\
\nabla ^{\prime }\Phi
\end{array}
\right) ,\left(
\begin{array}{c}
\nabla \Psi \\
\nabla ^{\prime }\Psi
\end{array}
\right) \right\rangle \right\rangle ,
\end{equation}

where

\begin{equation}
\label{2.19}
\begin{array}{c}
{\cal R}=\left(
\begin{array}{cc}
-a+\alpha & -\alpha -s \\
\alpha -s & -a-\alpha
\end{array}
\right) ,\alpha ^{*}=-\alpha ,a=\frac 12\left( r-r^{*}\right) ,s=\frac
12\left( r+r^{*}\right) , \\
\hbox{where}\ \alpha \in \stackrel{\circ }{\bf g}\wedge \stackrel{\circ }%
{\bf g}\text{ because }B\in End\left( \stackrel{\circ }{\bf g}\oplus
\stackrel{\circ }{\bf g}\right) ,
\end{array}
\end{equation}

and our choice of $B$ supposes that $s\in \stackrel{\circ }{\bf g}\otimes
\stackrel{\circ }{\bf g}$.The Jacobi identity for this bracket will be valid
if ${\cal R}$ satisfies the modified Yang-Baxter equation. In tensor
notations a Poisson brackets of monodromy matrix have the form:

\begin{equation}
\label{2.20}\left\{ M_1,M_2\right\} =\left( a-\alpha \right)
M_1M_2-M_1M_2\left( a+\alpha \right) +M_1\left( \alpha -s\right)
M_2+M_2\left( \alpha +s\right) M_1.
\end{equation}

The corresponding lattice Poisson algebra for which the monodromy matrix has
the brackets (\ref{2.20}) is:

\begin{equation}
\label{2.21}
\begin{array}{c}
\left\{ L_1^n,L_2^n\right\} =\left( a-\alpha \right)
L_1^nL_2^n-L_1^nL_2^n\left( a+\alpha \right) , \\
\left\{ L_1^n,L_2^{n+1}\right\} =L_1^n\left( \alpha -s\right) L_2^{n+1}, \\
\left\{ L_1^n,L_2^m\right\} =0,\left| n-m\right| \geq 2.
\end{array}
\end{equation}

Our next step is a classification of the Poisson brackets of type (\ref{2.9}%
) for which the Poisson brackets of the monodromy matrix may be regularized.
It is difficult to classify all non-skew solutions of the Yang-Baxter
equation for which there exists an $\alpha \in \stackrel{\circ }{\bf g}%
\wedge \stackrel{\circ }{\bf g}$ such that a matrix ${\cal R}$ in (\ref{2.19}%
) is a solution of the Yang-Baxter equation. But we can easily construct all
solutions of the Yang-Baxter equation for ${\bf g}\oplus {\bf g}$ according
to the Belavin-Drinfeld classification theorem \cite{bd}, and then choose
solutions of the form (\ref{2.19}). To realize this program we start with an
easy theorem:

\begin{theorem}
\label{theorem2.1} If ${\cal R}$ is a solution of the modified Yang-Baxter
equation for ${\bf g}\oplus {\bf g}$ of the type (\ref{2.19}), then $a+s$ is
a solution of the modified Yang-Baxter equation for ${\bf g}$.
\end{theorem}

{\it Proof}{\bf . }Let $r_1=-a+\alpha ,r_2=-\alpha -s$,then $r=-\left(
r_1+r_2\right) $. From the Yang-Baxter equation for ${\cal R}$ it follows:

\begin{equation}
\label{2.22}
\begin{array}{c}
\begin{array}{c}
\left[ r_1X,r_2Y\right] -r_1\left( \left[ r_1X,Y\right] +\left[
X,r_1Y\right] \right) =-\left[ X,Y\right] , \\
\left[ r_2X,r_2Y\right] -r_2\left( \left[ \left( r_1-2\alpha \right)
X,Y\right] +\left[ X,\left( r_1-2\alpha \right) Y\right] \right) =0,
\end{array}
\\
\left[ r_1X,r_2Y\right] -r_1\left[ X,r_2Y\right] -r_2\left[ \left(
r_2+2\alpha \right) X,Y\right] =0, \\
X,Y\in {\bf g}{\em .}
\end{array}
\end{equation}

It is easy to check that the (\ref{2.22}) implies the Yang-Baxter equation
for $r$:

\begin{equation}
\label{2.23}\left[ rX,rY\right] -r\left( \left[ rX,Y\right] +\left[
X,rY\right] \right) =-\left[ X,Y\right] .
\end{equation}

Let us turn now to the detailed study of the case of affine Lie algebras. We
use the terminology and notations of the book \cite{k}.

Let ${\bf g}$ be an affine Lie algebra, $\bigtriangleup _{+}$ the set of its
positive roots, $\stackrel{\circ }{\bf g}$ the corresponding spreading
simple Lie algebra, $\stackrel{\circ }{\bigtriangleup }_{+}$ the set of its
positive roots. Let $\bigtriangleup _{++}=\bigtriangleup _{+}\backslash
\stackrel{\circ }{\bigtriangleup }_{+}$. Using this notations we formulate
some version of the Belavin-Drinfeld classification theorem.

\begin{theorem}
\label{theorem2.2} \cite{bd} Up to an automorphism any solution of the
modified classical Yang-Baxter equation for an affine Lie algebra ${\bf g}$
has the form:%
$$
R=\sum\limits_{\alpha \in \bigtriangleup _{++}}e_\alpha \wedge e_{-\alpha
}+r,
$$
where $r$ is a solution of the modified Yang-Baxter equation for $\stackrel{%
\circ }{\bf g}$.
\end{theorem}

For an explicit form of such solutions see \cite{bd}.

In \cite{bd} such solutions are called trigonometric.

Thus from theorem 2.2 we have the following ansatz for $a$:

\begin{equation}
\label{2.25}a=\sum\limits_{\alpha \in \bigtriangleup _{++}}e_\alpha \wedge
e_{-\alpha }+a_0,a_0\in \stackrel{\circ }{\bf g}\wedge \stackrel{\circ }{\bf %
g}.
\end{equation}

and we reduce our problem to the Yang-Baxter equation on the square of a
finite dimensional Lie algebra $\stackrel{\circ }{\bf g}$. Namely, ${\cal R}$
is a solution of the Yang-Baxter equation iff

\begin{equation}
\label{2.26}\left(
\begin{array}{cc}
-a_0+\alpha & -\alpha -s \\
\alpha -s & -a_0-\alpha
\end{array}
\right)
\end{equation}

is a solution of the Yang-Baxter equation for $\stackrel{\circ }{\bf g}%
\oplus \stackrel{\circ }{\bf g}$.

\begin{theorem}
\label{theorem2.3} Let%
$$
\left(
\begin{array}{cc}
A & B \\
B^{*} & D
\end{array}
\right) ,A^{*}=-A,D^{*}=-D
$$
be a solution of the Yang-Baxter equation for $\stackrel{\circ }{\bf g}%
\oplus \stackrel{\circ }{\bf g}$. It has the form (\ref{2.26}) iff
\begin{equation}
\label{2.28}A+B=B^{*}+D.
\end{equation}
Under this condition
\end{theorem}

\begin{equation}
\label{2.29}\alpha =\frac{B^{*}-B}2,\ a_0=-\frac{A+D}2,\ s=-\frac{B+B^{*}}2.
\end{equation}

Notice that we again come to the condition (\ref{1.3}).

{\it Remark}. It may be showed that it is not necessary to impose the
condition $B\in End\left( \stackrel{\circ }{\bf g}\oplus \stackrel{\circ }%
{\bf g}\right) $ a priori. Actually, this condition follows from the
generalization of theorem 2.2 for a direct sum of two copies of an affine
Lie algebra, because $\alpha \in \stackrel{\circ }{\bf g}\wedge \stackrel{%
\circ }{\bf g},\ s\in \stackrel{\circ }{\bf g}\otimes \stackrel{\circ }{\bf g%
}$ for every solution of the modified Yang-Baxter equation for ${\bf g}%
\oplus {\bf g}$ of the form (\ref{2.19}).

Unfortunately, condition (\ref{2.28}) is not stable under the automorphisms
of $\stackrel{\circ }{\bf g}$. So we cannot use the Belavin-Drinfeld theorem
to classify all regularizations. But this theorem gives a possibility to
construct sufficiently general examples of such regularizations. These
examples will be presented in the next section. In the $\widehat{sl\left(
2\right) }$ case we shall able to classify all regularizations.

\section{Main examples of regularizations}

Now we are ready to discuss examples of regularizations using the results of
the previous sections. We shall consider affine Lie algebras of type $%
X_N^{\left( 1\right) }$ and $X_N^{\left( 2\right) }$ in the loop
realization. We shall describe the corresponding lattice quantum algebras
and their Casimir elements. In the $\widehat{sl\left( 2\right) }$ case we
shall explain the Algebraic Bethe Ansatz construction for such algebras.

\begin{example}
{\bf Nontwisted loop algebras.}
\end{example}

The first example is connected with the r-matrix of the double of a finite
dimensional Lie algebra $\stackrel{\circ }{\bf g}$ equipped with the
structure of a quasitriangular Lie bialgebra. Let ${\bf g}$ be an affine Lie
algebra of type $X_N^{\left( 1\right) }$, $\stackrel{\circ }{\bf g}$ the
corresponding finite-dimensional Lie algebra. To apply theorem 2.3 consider
the r-matrix of its double $\stackrel{\circ }{\bf d}=\stackrel{\circ }{\bf g}%
\oplus \stackrel{\circ }{\bf g};$ we have
\begin{equation}
\label{3.1}r=\left(
\begin{array}{cc}
\alpha & -2\alpha _{+} \\
2\alpha _{-} & -\alpha
\end{array}
\right) ,
\end{equation}
where $\alpha $ is some solution of the modified Yang-Baxter equation for $%
\stackrel{\circ }{\bf g}$ and $\alpha _{\pm }=\frac 12\left( \alpha \pm
I\right) $. ($I$ is the identity operator in $\stackrel{\circ }{\bf g}$; its
kernel is the Casimir element $t.$) According to theorem 2.3, in this case
one gets:

\begin{equation}
\label{3.2}s=I,a_0=0.
\end{equation}

In this case $r=a+I$ is the rational r-matrix for ${\bf g}$. We choose for $%
{\bf g}$ the non-twisted loop realization \cite{k}. We remind that in this
realization ${\bf g}=\stackrel{\circ }{\bf g}\otimes {\bf C}\left[ \lambda
,\lambda ^{-1}\right] $, and the invariant bilinear form is given by

\begin{equation}
\label{3.3}\left\langle X\left( \lambda \right) ,Y\left( \lambda \right)
\right\rangle =Res\text{ }tr\left( X\left( \lambda \right) Y\left( \lambda
\right) \right) \frac{d\lambda }\lambda ,
\end{equation}

where $tr$ is an invariant bilinear form on $\stackrel{\circ }{\bf g}$.

The kernel of $a$ in this realization is:

\begin{equation}
\label{3.4}a\left ( \lambda ,\mu \right) =-t\frac{\lambda +\mu }{\lambda
-\mu },
\end{equation}

where we identify ${\bf g}\otimes {\bf g}$ with $\stackrel{\circ }{\bf g}%
\otimes \stackrel{\circ }{\bf g}\otimes {\bf C}\left[ \lambda ,\lambda
^{-1}\right] \otimes {\bf C}\left[ \mu ,\mu ^{-1}\right] $. Thus we have the
following formulas for $-a\pm \alpha $:

\begin{equation}
\label{3.5}
\begin{array}{c}
-a+\alpha =\frac \lambda {\lambda -\mu }2\alpha _{+}-\frac \mu {\lambda -\mu
}2\alpha _{-}, \\
-a-\alpha =\frac \mu {\lambda -\mu }2\alpha _{+}-\frac \lambda {\lambda -\mu
}2\alpha _{-}.
\end{array}
\end{equation}

Let ${\cal R}_{\pm }$ be the finite-dimensional quantum R-matrix in the
fundamental representation, which corresponds to $2\alpha _{\pm }$ after
quantization. In the classical limit

\begin{equation}
\label{3.6}{\cal R}_{\pm }=I+2\alpha _{\pm }h+o\left( h\right) ,
\end{equation}

where $h$ is the deformation parameter.

We have:

\begin{equation}
\label{3.7}{\cal R}_{-}=P\left( {\cal R}_{+}^{-1}\right) ,
\end{equation}

where $P$ is the permutation operator in the tensor square.

Using these data we may construct the quantum R-matrices corresponding to $%
-a\pm \alpha $. If we denote the quantum R-matrix corresponding to $%
-a-\alpha $ by ${\cal R}\left( \lambda ,\mu \right) ,$ then

\begin{equation}
\label{3.8}{\cal R}\left( \lambda ,\mu \right) =\frac \lambda {\lambda -\mu }%
{\cal R}_{-}^{-1}-\frac \mu {\lambda -\mu }{\cal R}_{+}^{-1},
\end{equation}

and the quantum R-matrix ${\cal R}\left( \lambda ,\mu \right) ^T$
corresponds to $-a+\alpha $:

\begin{equation}
\label{3.9}{\cal R}\left( \lambda ,\mu \right) ^T=\frac \lambda {\lambda
-\mu }{\cal R}_{+}-\frac \mu {\lambda -\mu }{\cal R}_{-},
\end{equation}

here $T$ is the conjugation with respect to the scalar product $tr$.

Finally, we have the quantum R-matrix on the square of ${\bf g:}$

\begin{equation}
\label{3.10}{\cal R}_q=\left(
\begin{array}{cc}
{\cal R}\left( \lambda ,\mu \right) ^T & {\cal R}_{+}^{-1} \\ {\cal R}_{-} &
{\cal R}\left( \lambda ,\mu \right)
\end{array}
\right) .
\end{equation}

According to theorem 1.2 one can get the relations in the quantum lattice
algebra $Fun_q^{{\cal R}}\left( {\sf G}_\tau \right) $, which gives a
lattice approximation of the continuous system:

\begin{equation}
\label{3.11}
\begin{array}{c}
{\cal R}\left( \lambda ,\mu \right) ^TL_1^n\left( \lambda \right)
L_2^n\left( \mu \right) =L_2^n\left( \mu \right) L_1^n\left( \lambda \right)
{\cal R}\left( \lambda ,\mu \right) , \\ L_1^n\left( \lambda \right)
{\cal R}_{-}^{-1}L_2^{n+1}\left( \mu \right) =L_2^{n+1}\left( \mu \right)
L_1^n\left( \lambda \right) , \\ L^n\left( \lambda \right) \in Fun_q^{{\cal R%
}}\left( {\sf G}_\tau \right) \otimes End\left( V\right) ,
\end{array}
\end{equation}

here $V$ is the fundamental representation space.

 From theorem 1.3 we get the following commutation relations for the
monodromy matrix:

\begin{equation}
\label{3.12}{\cal R}\left( \lambda ,\mu \right) ^TM_1\left( \lambda \right)
{\cal R}_{-}^{-1}M_2\left( \mu \right) =M_2\left( \mu \right) {\cal R}%
_{+}M_1\left( \lambda \right) {\cal R}\left( \lambda ,\mu \right).
\end{equation}

The algebra (\ref{3.11}) is connected with the Lattice Kac-Moody algebra $%
{\cal A}_{LC}$ \cite{afs}. Namely, the algebra (\ref{3.11}) admits a family
of representation for which there exists the limit

\begin{equation}
\label{3.13}L^n\left( \lambda \right) \stackunder{\lambda \rightarrow 0}{%
\longrightarrow }\lambda ^{-k_n}\stackrel{\circ }{L}^n+o\left( \lambda
^{-k_n}\right) ,{k_n}\in {\bf Z}.
\end{equation}
{\em \ }

 From (\ref{3.8}), (\ref{3.9}) we have the asymptotic conditions for ${\cal R}%
\left( \lambda ,\mu \right) ,{\cal R}\left( \lambda ,\mu \right) ^T$:

\begin{equation}
\label{3.14}
\begin{array}{c}
{\cal R}\left( \lambda ,\mu \right) \stackunder{\mu \rightarrow 0}{%
\longrightarrow }{\cal R}_{-}^{-1}, \\ {\cal R}\left( \lambda ,\mu \right) ^T%
\stackunder{\mu \rightarrow 0}{\longrightarrow }{\cal R}_{+}.
\end{array}
\end{equation}

Using (\ref{3.11}),(\ref{3.14}), one can get the relations for $\stackrel{%
\circ }{L}^n$:

\begin{equation}
\label{3.15}
\begin{array}{c}
\stackrel{\circ }{L}_2^n\stackrel{\circ }{L}_1^n={\cal R}_{+}\stackrel{\circ
}{L}_1^n\stackrel{\circ }{L}_2^n{\cal R}_{-} \\ \stackrel{\circ }{L}_1^n%
{\cal R}_{-}^{-1}\stackrel{\circ }{L}_2^{n+1}=\stackrel{\circ }{L}_2^{n+1}%
\stackrel{\circ }{L}_1^n
\end{array}
\end{equation}

These are the relations in the Lattice Kac-Moody algebra ${\cal A}_{LC}$ .
If $\alpha =\sum\limits_{\alpha \in \stackrel{\circ }{\bigtriangleup }%
_{+}}e_\alpha \wedge e_{-\alpha }$, then the monodromy matrix for this
algebra

\begin{equation}
\label{3.16}\stackrel{\circ }{M}=\stackrel{\circ }{L}^1\cdot \ldots \cdot
\stackrel{\circ }{L}^N
\end{equation}

satisfies the commutation relations for the quantum group $U_q\left(
\stackrel{\circ }{\bf g}\right) $:

\begin{equation}
\label{3.17}{\cal R}_{+}\stackrel{\circ }{M}_1{\cal R}_{-}^{-1}\stackrel{%
\circ }{M}_2=\stackrel{\circ }{M}_2{\cal R}_{+}\stackrel{\circ }{M}_1{\cal R}%
_{-}^{-1}.
\end{equation}

This construction is useful for the computation of the center of the algebra
(\ref{3.11}).

\begin{theorem}
\label{theorem3.1} For generic $q$ and $N$ odd the center of the algebra (%
\ref{3.11}) contains the elements
\begin{equation}
\label{3.18}C_k=tr_q\left( \stackrel{\circ }{M}\right) ^k=tr\text{ }q^{2\rho}%
\left( \stackrel{\circ }{M}\right) ^k,k=1,\ldots ,rk\stackrel{\circ }{\bf g}.
\end{equation}
(Here $\rho $ is half the sum of positive roots in $U_q\left( \stackrel{%
\circ }{\bf g}\right) $.)
\end{theorem}

It is natural to expect that in an appropriate topology these elements
generate the center of the algebra (\ref{3.11}).

The proof of this theorem is complete similar to the proof of such theorem
for the algebra ${\cal A}_{LC}$ \cite{afs}.

 From this theorem one can deduce that the center of the algebra (\ref{3.11})
coincides with the center of ${\cal A}_{LC}$ \cite{afs} and with the center
of $U_q\left( \stackrel{\circ }{\bf g}\right) $. Thus we have

\begin{corollary}
The extensions
$$
centU_q\left( \stackrel{\circ }{\bf g}\right) \subset U_q\left( \stackrel{%
\circ }{\bf g}\right) \subset {\cal A}_{LC}\subset Fun_q^{{\cal R}}\left(
{\sf G}_\tau \right)
$$
are central.
\end{corollary}

\begin{example}
{\bf Twisted loop algebras.}
\end{example}

We are leaving for a short time our first example to consider the second
one. Let us consider an affine Lie algebra ${\bf g}$ of a type $X_N^{\left(
r\right) },r=1,2,3$. Let
\begin{equation}
\label{3.20}a_0=\sum\limits_{\alpha \in \stackrel{\circ }{\bigtriangleup }%
_{+}}e_\alpha \wedge e_{-\alpha },\alpha =0,s\in End\left( \stackrel{\circ }%
{\bf h}\right) .
\end{equation}

It is not difficult to verify that the corresponding r-matrix of the type (%
\ref{2.26}) satisfies the Yang-Baxter equation for $\stackrel{\circ }{\bf g}%
\oplus \stackrel{\circ }{\bf g}$. In this case $r=P_{+}-P_{-}+s$, where $%
P_{\pm }$ are projection operators onto the opposite Borel subalgebras of $%
{\bf g}$. We shall use a twisted loop realization for ${\bf g}$ with
invariant bilinear form (\ref{3.3}) \cite{k}, \cite{bd}. In this realization
${\bf g}$ is the set of stable points of an automorphism of an affine Lie
algebra $X_{N^{\prime }}^{\left( 1\right) }$ in a non-twisted loop
realization. Here $N^{\prime }$ does not in general coincide with $N$. The
automorphism $\widehat{C}$ of $X_{N^{\prime }}^{\left( 1\right) }$ is given
by

\begin{equation}
\label{3.21}\widehat{C}X\left( \lambda \right) =CX\left( \lambda e^{-\frac{%
2\pi i}{rh}}\right) ,
\end{equation}

where $h$ is the Coxeter number of the algebra $X_N^{\left( r\right) }$, $C$
is the Coxeter automorphism of $\stackrel{\circ }{\bf g}^{\prime }$ which
corresponds to the affine Lie algebra $X_{N^{\prime }}^{\left( 1\right) }$.

If ${\bf G}=X_N^{\left( 1\right) }$, $N^{\prime }$ coincides with $N$ and $%
\stackrel{\circ }{\bf g}^{\prime }$ coincides with $\stackrel{\circ }{\bf g}$.
For simplicity we restrict ourselves to this case. The automorphism $C$
has degree $h$ and the algebra $\stackrel{\circ }{\bf g}$ is decomposed into
the direct sum $\stackrel{\circ }{\bf g}=\bigoplus\limits_{j=0}^{h-1}%
\stackrel{\circ }{\bf g}_j,\stackrel{\circ }{\bf g}_0=\stackrel{\circ }{\bf h%
}$, where $\stackrel{\circ }{\bf g}_j$ is the eigenspace of $C$
corresponding to the eigenvalue $e^{\frac{2\pi i}hj}$. The Casimir element
of $\stackrel{\circ }{\bf g}$ is decomposed into the sum $%
t=\sum\limits_{j=0}^{h-1}t_j,t_j\in \stackrel{\circ }{\bf g}_j\otimes
\stackrel{\circ }{\bf g}_{-j},t_0\in \stackrel{\circ }{\bf h}\otimes $ $%
\stackrel{\circ }{\bf h}$ . The kernel of the operator $a$ in this
realization is:

\begin{equation}
\label{3.22}a\left( \lambda ,\mu \right) =-\frac{\sum\limits_{j=0}^{h-1}t_j%
\left( \lambda ^{h-j}\mu ^j+\lambda ^j\mu ^{h-j}\right) }{\lambda ^h-\mu ^h}.
\end{equation}

We suppose that $s=t_0$. Now we are ready to describe the quantization. Let $%
{\cal R}\left( \lambda ,\mu \right) $ be the quantum R-matrix in the
fundamental representation $V$ which corresponds to $a\left( \lambda ,\mu
\right) $ after quantization. Evidently, such quantum r-matrix exists
because the classical r-matrix $a$ coincides with the classical r-matrix $%
a+\alpha $ from example 1 if $\alpha =\sum\limits_{\alpha \in \stackrel{%
\circ }{\bigtriangleup }_{+}}e_\alpha \wedge e_{-\alpha }$. For the latter
classical r-matrix the corresponding quantum r-matrix exists (see example
1). Let ${\cal R}_0=e^{hs}$. We have the following quantum R-matrix on the
square:

\begin{equation}
\label{3.23}{\cal R}_q=\left(
\begin{array}{cc}
{\cal R}\left( \lambda ,\mu \right) ^{-1} & {\cal R}_0^{-1} \\ {\cal R}%
_0^{-1} & {\cal R}\left( \lambda ,\mu \right) ^{-1}
\end{array}
\right) .
\end{equation}

In the standard way we built the quantum lattice algebra $Fun_q^{{\cal R}%
}\left( {\sf G}_\tau \right) $ which gives a lattice counterpart for our
continuous system. According to theorem 1.2, the commutation relations for
it have the form
\begin{equation}
\label{3.24}
\begin{array}{c}
{\cal R}\left( \lambda ,\mu \right) L_1^n\left( \lambda \right) L_2^n\left(
\mu \right) =L_2^n\left( \mu \right) L_1^n\left( \lambda \right) {\cal R}%
\left( \lambda ,\mu \right) , \\ L_1^n\left( \lambda \right)
{\cal R}_0L_2^{n+1}\left( \mu \right) =L_2^{n+1}\left( \mu \right)
L_1^n\left( \lambda \right) , \\ L^n\left( \lambda \right) \in Fun_q^{{\cal R%
}}\left( {\sf G}_\tau \right) \otimes End\left( V\right) .
\end{array}
\end{equation}

For the monodromy matrix we have the relation:

\begin{equation}
\label{3.25}{\cal R}\left( \lambda ,\mu \right) M_1\left( \lambda \right)
{\cal R}_0M_2\left( \mu \right) =M_2\left( \mu \right) {\cal R}_0M_1\left(
\lambda \right) {\cal R}\left( \lambda ,\mu \right) .
\end{equation}

Similarly to theorem 3.1, there exists a description of the center of the
algebra (\ref{3.24}).

\begin{theorem}
\label{theorem3.2} Let
\begin{equation}
\label{3.26}M\stackunder{\lambda \rightarrow 0}{\longrightarrow }\lambda
^{-k}\stackrel{\circ }{M}+o\left( \lambda ^{-k}\right) .
\end{equation}
For a generic $q$ and $N$ odd the center of the algebra (\ref{3.24})
contains the elements:
\begin{equation}
\label{3.27}C_k=tr\left( \stackrel{\circ }{M}\right) ^k,k=1,\ldots ,rk%
\stackrel{\circ }{\bf g}.
\end{equation}
\end{theorem}

As in theorem 3.1, it is natural to expect that in an appropriate
topology these elements generate the center of the algebra (\ref{3.24}).

The algebra (\ref{3.24}), as opposed to the algebra (\ref{3.11}), possesses
a remarkable property; namely, it admits an ultralocalization. This property
is important for the construction of representations of lattice algebras.
The exact statement is given by the following theorem.

\begin{theorem}
Let $\rho $ be the fundamental representation of the group $\stackrel{\circ
}{H}$ corresponding to the algebra $\stackrel{\circ }{\bf h}$, and $V$ the
space of its fundamental representation used for the definition of the
algebra (\ref{3.24}).Then $\rho $ acts naturally in $V$. Let $A$ be the free
algebra generated by the matrix coefficients of the matrices $\widehat{L}%
^n\in A\otimes EndV,G^n\in A\otimes \rho $, satisfying the relations:
\begin{equation}
\label{3.29}
\begin{array}{c}
{\cal R}\left( \lambda ,\mu \right) \widehat{L}_1^n\left( \lambda \right)
\widehat{L}_2^n\left( \mu \right) =\widehat{L}_2^n\left( \mu \right)
\widehat{L}_1^n\left( \lambda \right) {\cal R}\left( \lambda ,\mu \right) ,
\\ G_1^n
\widehat{L}_2^{n+1}\left( \mu \right) ={\cal R}_0^{\frac 12}\widehat{L}%
_2^{n+1}\left( \mu \right) G_1^n, \\ G_1^n\widehat{L}_2^n\left( \mu \right) =%
\widehat{L}_2^n\left( \mu \right) G_1^n{\cal R}_0^{\frac 12}.
\end{array}
\end{equation}
{\em \ }

Then there exist the homomorphism of the algebras:
\begin{equation}
\label{3.30}
\begin{array}{c}
Fun_q^{
{\cal R}}\left( {\sf G}_\tau \right) \rightarrow A, \\ L^n\mapsto G^{n-1}
\widehat{L}^nG^{n^{-1}}, \\ M\mapsto G^N
\widehat{M}G^{N^{-1}}; \\ \text{here }\widehat{M}=\widehat{L}^1\cdot \ldots
\cdot \widehat{L}^N.
\end{array}
\end{equation}
\end{theorem}

{\it Remark. }This homomorphism has the form of a lattice gauge
transformation: the generators $L^n$ of the nonultralocal algebra $Fun_q^{%
{\cal R}}\left( {\sf G}_\tau \right) $ are connected by a lattice gauge
transformation with the generators $\widehat{L}^n$ satisfying ultralocal
relations. Therefore this homomorphism is called ultralocalization. The
monodromy matrices $M$ and $\widehat{M}$ are conjugate and hence
nonultralocality is equivalent to including some quasiperiodic boundary
conditions. This is natural in the spirit of the operator extension theory
used in the previous section. In particular examples it is easier to
construct representations of the algebra $A$ than those of the algebra (\ref
{3.24}). It is the main motivation for the definition of the algebra $A$.

The proof of this theorem consists in a straightforward check of the
relations (\ref{3.24}) for the generators $G^{n-1}\widehat{L}^nG^{n^{-1}}$ .
A similar theorem on ultralocalization exists for the algebra ${\cal A}_{LC}$
\cite{fv2}.

Now we are ready to describe all regularized r-matrices and their
regularizations for the algebra $A_1^{\left( 1\right) }$. According to the
Belavin-Drinfeld theorem for finite-dimensional semisimple Lie algebras \cite
{bd}, we may construct all solutions of the modified Yang-Baxter equation on
the square of corresponding finite-dimensional Lie algebra $\stackrel{\circ }%
{\bf g}=sl\left( 2\right) $. Modulo some freedom in the choice of the Cartan
components of R-matrix on the skew-diagonal there exist only two solutions
of the Yang-Baxter equation on $sl\left( 2\right) \oplus sl\left( 2\right) $
considered in examples 1 and 2 in this section. Moreover, the r-matrix of
the double (\ref{3.1}) in example 1 is connected with the unique structure
of a bialgebra on $sl\left( 2\right) $ for which

\begin{equation}
\label{3.31}\alpha =\sum\limits_{\alpha \in \stackrel{\circ }{\bigtriangleup
}_{+}}e_\alpha \wedge e_{-\alpha }.
\end{equation}

For the algebra $A_1^{\left( 1\right) }$ the Coxeter number $h$ equals 2,
and the Coxeter automorphism has the form:

\begin{equation}
\label{3.32}
\begin{array}{c}
CX=DXD^{-1}, \\
D=\left(
\begin{array}{cc}
1 & 0 \\
0 & -1
\end{array}
\right) .
\end{array}
\end{equation}

We summarize all results for the algebra $A_1^{\left( 1\right) }$ in the
following theorem. We formulate at once the quantum version of all formulas.
The reader may reproduce the corresponding classical formulas by a
quasiclassical limit.

\begin{theorem}
For the algebra $A_1^{\left( 1\right) }$ there exists only two nonultralocal
Lie-Poisson brackets of the type (\ref{2.9}) which may be regularized in the
way (\ref{2.15}). Namely, let $b_{\pm }$ be the opposite Borel subalgebras
in $A_1^{\left( 1\right) }$ and $\stackrel{\circ }{b_{\pm }}$ the opposite
Borel subalgebras in $sl\left( 2\right) $. Then the regularized r-matrices
and the corresponding lattices algebras are:

\begin{enumerate}
\item  $r=P_{b_{+}\backslash \stackrel{\circ }{b_{+}}}-P_{b_{-}\backslash
\stackrel{\circ }{b_{-}}}+I$ is the regularized r-matrix, here $%
P_{b_{+}\backslash \stackrel{\circ }{b_{+}}},P_{b_{-}\backslash \stackrel{%
\circ }{b_{-}}}$ are projectors onto the corresponding subspaces, $I$ is the
identity operator in $sl\left( 2\right) $. It is precisely the rational
r-matrix. If we use the non-twisted current realization $A_1^{\left(
1\right) }=sl\left( 2\right) \otimes {\bf C}\left[ \lambda ,\lambda
^{-1}\right] $ and the scalar product (\ref{3.3}), its kernel is given by

\begin{equation}
\label{3.33}r\left( \lambda ,\mu \right) =-\frac{\mu t}{\lambda -\mu },
\end{equation}

where $t$ is the Casimir element for $sl\left( 2\right) $.

The relations in the corresponding lattice algebra are given by formulas (%
\ref{3.11}),where for the fundamental representation $A_1^{\left( 1\right) }$
in ${\bf C}^2\left[ \lambda ,\lambda ^{-1}\right] $:

\begin{equation}
\label{3.34}
\begin{array}{c}
{\cal R}_{+}=q^{-\frac 12}\left(
\begin{array}{cccc}
q & 0 & 0 & 0 \\
0 & 1 & q-q^{-1} & 0 \\
0 & 0 & 1 & 0 \\
0 & 0 & 0 & q
\end{array}
\right) , \\
{\cal R}_{-}=P\left( {\cal R}_{+}^{-1}\right) ,
\end{array}
\end{equation}

\begin{equation}
\label{3.35}{\cal R}\left( \lambda ,\mu \right) =\left(
\begin{array}{cccc}
\frac{\lambda q-\mu q^{-1}}{\lambda -\mu } & 0 & 0 & 0 \\
0 & 1 & \frac \mu {\lambda -\mu }\left( q-q^{-1}\right)  & 0 \\
0 & \frac \lambda {\lambda -\mu }\left( q-q^{-1}\right)  & 1 & 0 \\
0 & 0 & 0 & \frac{\lambda q-\mu q^{-1}}{\lambda -\mu }
\end{array}
\right)
\end{equation}
$.$

\item  $r=P_{b_{+}}-P_{b_{-}}+\xi P_{\stackrel{\circ }{\em h}}$ is the
regularized r-matrix, $P_{\stackrel{\circ }{\em h}}$ is the projection
operator  onto the Cartan subalgebra of $sl\left( 2\right) $. For simplicity
we consider only the case $\xi =1$. Fix the twisted current realization

\begin{equation}
\label{3.36}A_1^{\left( 1\right) }=\left\{ X\left( \lambda \right) \in
sl\left( 2\right) \otimes {\bf C}\left[ \lambda ,\lambda ^{-1}\right]
:X\left( -\lambda \right) =DX\left( \lambda \right) D^{-1}\right\}
\end{equation}

and  the scalar product (\ref{3.3}). Then the kernel of $r$ is given by

\begin{equation}
\label{3.37}r\left( \lambda ,\mu \right) =-\frac{\mu ^2t_0}{\lambda ^2-\mu ^2%
}-\frac{\lambda \mu t_1}{\lambda ^2-\mu ^2},
\end{equation}

where $t_0$ is the $\stackrel{\circ }{\bf h}-$ component of the $sl\left(
2\right) $ Casimir element and $t_1\in \stackrel{\circ }{\bf g}_1\otimes
\stackrel{\circ }{\bf g}_{-1},\stackrel{\circ }{\bf g}_1\in \left\{ X\in
sl\left( 2\right) ,DXD^{-1}=-X\right\} $. The relations in the corresponding
lattice algebra are given by the formulas (\ref{3.24}), where for the
fundamental representation $A_1^{\left( 1\right) }$ in ${\bf C}^2\left[
\lambda ,\lambda ^{-1}\right] $ :

\begin{equation}
\label{3.38}{\cal R}_0=q^{-\frac 12}\left(
\begin{array}{cccc}
q & 0 & 0 & 0 \\
0 & 1 & 0 & 0 \\
0 & 0 & 1 & 0 \\
0 & 0 & 0 & q
\end{array}
\right) ,
\end{equation}

\begin{equation}
\label{3.39}{\cal R}\left( \lambda ,\mu \right) =\left(
\begin{array}{cccc}
\frac{\lambda ^2q-\mu ^2q^{-1}}{\lambda ^2-\mu ^2} & 0 & 0 & 0 \\
0 & 1 & \frac{2\lambda \mu }{\lambda ^2-\mu ^2}\left( q-q^{-1}\right)  & 0
\\
0 & \frac{2\lambda \mu }{\lambda ^2-\mu ^2}\left( q-q^{-1}\right)  & 1 & 0
\\
0 & 0 & 0 & \frac{\lambda ^2q-\mu ^2q^{-1}}{\lambda ^2-\mu ^2}
\end{array}
\right) .
\end{equation}
\end{enumerate}
\end{theorem}

{\it Remark.}

The quantum R-matrices (\ref{3.35}) and (\ref{3.39}) are in fact the
ordinary trigonometric quantum R-matrices in different realizations. To see
this just notice that the classical r-matrices $a+\alpha $ from the example
1 (see (\ref{3.2}),(\ref{2.25}),(\ref{3.31})) and $a$ from the example 2
(see (\ref{2.25}), (\ref{3.20}) coincide.

We conclude this section with the Bethe-Ansatz construction for algebras
with relations (\ref{3.11}), (\ref{3.24}) in the $A_1^{\left( 1\right) }$
case.

We start with example 1.The 'generating function' producing an infinite
series of quantum commuting integrals of motion \cite{f2}, \cite{f3} is

\begin{equation}
\label{3.40}
\begin{array}{c}
tr_qM\left( \lambda \right) =qA\left( \lambda \right) +q^{-1}D\left( \lambda
\right) , \\
\text{where }M\left( \lambda \right) =\left(
\begin{array}{cc}
A\left( \lambda \right) & B\left( \lambda \right) \\
C\left( \lambda \right) & D\left( \lambda \right)
\end{array}
\right) \\
\text{and }\left[ tr_qM\left( \lambda \right) ,tr_qM\left( \mu \right)
\right] =0.
\end{array}
\end{equation}

We suppose that there exists a reference state $\Omega $:

\begin{equation}
\label{3.41}A\left( \lambda \right) \Omega =a\left( \lambda \right) \Omega
,D\left( \lambda \right) \Omega =d\left( \lambda \right) \Omega ,C\left(
\lambda \right) \Omega =0,B\left( \lambda \right) \Omega \neq 0.
\end{equation}

We shall try to seek a representation of the algebra (\ref{3.12}) in which $%
tr_qM\left( \lambda \right) $ is diagonal. According to the standard Bethe
Ansatz technique we try to find the eigenvectors of $tr_qM\left( \lambda
\right) $ in the form:

\begin{equation}
\label{3.42}\psi \left( \lambda _1,\ldots ,\lambda _n\right) =B\left(
\lambda _1\right) \ldots B\left( \lambda _n\right) \Omega .
\end{equation}

The relations between $A\left( \lambda \right) ,B\left( \lambda \right)
,C\left( \lambda \right) ,D\left( \lambda \right) $ which are essential to
calculate the spectrum of $tr_qM\left( \lambda \right) $ are:

\begin{equation}
\label{3.43}
\begin{array}{c}
\begin{array}{c}
\left[ B\left( \lambda \right) ,B\left( \mu \right) \right] =0, \\
A\left( \lambda \right) B\left( \mu \right) =
\frac{\mu -\lambda q^{-2}}{\mu -\lambda }B\left( \mu \right) A\left( \lambda
\right) -\frac \lambda {\mu -\lambda }\left( 1-q^{-2}\right) B\left( \lambda
\right) A\left( \mu \right) - \\ -\left( 1-q^{-2}\right) B\left( \lambda
\right) D\left( \mu \right) ,
\end{array}
\\
D\left( \lambda \right) B\left( \mu \right) =\frac \mu {\mu -\lambda }\left(
q^2-1\right) B\left( \lambda \right) D\left( \mu \right) +\frac{\mu -\lambda
q^2}{\mu -\lambda }B\left( \mu \right) D\left( \lambda \right) .
\end{array}
\end{equation}

The vector $\psi \left( \lambda _1,\ldots ,\lambda _n\right) $ will be an
eigenvector of $tr_qM\left( \lambda \right) $ with the eigenvalue

\begin{equation}
\label{3.44}qa\left( \lambda \right) \prod\limits_{i=1}^n\frac{\lambda
q^{-2}-\lambda _i}{\lambda -\lambda _i}+q^{-1}d\left( \lambda \right)
\prod\limits_{i=1}^n\frac{\lambda q^2-\lambda _i}{\lambda -\lambda _i}
\end{equation}

iff $\lambda _i$ satisfy the equations \cite{f2},\cite{f3}:

\begin{equation}
\label{3.45}\frac{d\left( \lambda _j\right) }{a\left( \lambda _j\right) }%
=\prod\limits_{i\neq j}\frac{\lambda _jq^{-2}-\lambda _i}{\lambda
_jq^2-\lambda _i}.
\end{equation}

Similarly, in example 2 the element

\begin{equation}
\label{3.46}
\begin{array}{c}
trM\left( \lambda \right) =A\left( \lambda \right) +D\left( \lambda \right)
, \\
\text{where }M\left( \lambda \right) =\left(
\begin{array}{cc}
A\left( \lambda \right) & B\left( \lambda \right) \\
C\left( \lambda \right) & D\left( \lambda \right)
\end{array}
\right)
\end{array}
\end{equation}

produces an infinite series of commuting conservation laws. Under the
assumption (\ref{3.41}) we look for the eigenvectors of $trM\left( \lambda
\right) $ which have the form (\ref{3.42}). The essential commutation
relations for this procedure are (\ref{3.25}):

\begin{equation}
\label{3.47}
\begin{array}{c}
\begin{array}{c}
\left[ B\left( \lambda \right) ,B\left( \mu \right) \right] =0, \\
A\left( \lambda \right) B\left( \mu \right) =\frac{\mu ^2-\lambda ^2q^{-2}}{%
\mu ^2-\lambda ^2}B\left( \mu \right) A\left( \lambda \right) -\frac{%
2\lambda \mu }{\mu ^2-\lambda ^2}\left( 1-q^{-2}\right) B\left( \lambda
\right) A\left( \mu \right) ,
\end{array}
\\
D\left( \lambda \right) B\left( \mu \right) =\frac{2\lambda \mu }{\mu
^2-\lambda ^2}\left( q^2-1\right) B\left( \lambda \right) D\left( \mu
\right) +\frac{\mu ^2-\lambda ^2q^2}{\mu ^2-\lambda ^2}B\left( \mu \right)
D\left( \lambda \right) .
\end{array}
\end{equation}

The vector $\psi \left( \lambda _1,\ldots ,\lambda _n\right) $ is an
eigenvector of $trM\left( \lambda \right) $ with the eigenvalue

\begin{equation}
\label{3.48}a\left( \lambda \right) \prod\limits_{i=1}^n\frac{\lambda
^2q^{-2}-\lambda _i^2}{\lambda ^2-\lambda _i^2}+d\left( \lambda \right)
\prod\limits_{i=1}^n\frac{\lambda ^2q^2-\lambda _i^2}{\lambda ^2-\lambda _i^2%
}
\end{equation}

iff $\lambda _{i\text{ }}$ satisfy the equations:

\begin{equation}
\label{3.49}q^2\frac{d\left( \lambda _j\right) }{a\left( \lambda _j\right) }%
=\prod\limits_{i\neq j}\frac{\lambda _j^2q^{-2}-\lambda _i^2}{\lambda
_j^2q^2-\lambda _i^2}.
\end{equation}

We omit the standard Bethe Ansatz calculations leading to these formulas (%
\ref{3.44}), (\ref{3.45}), (\ref{3.48}), (\ref{3.49}) \cite{stf}. It is easy
to verify that the modification of the commutation relations (\ref{3.43}), (%
\ref{3.47}) (as compared to the standard XXZ relations) only weakly
influences the Bethe Ansatz construction, so that the principal idea \cite
{stf} may be applied as before.

\section{Application to the nonlinear sigma model}

We recall some facts about chiral fields with values in Riemannian symmetric
spaces. Let $\left( {\bf k}{\em ,}{\bf p}\right) $ be a Riemannian symmetric
pair for a semisimple finite-dimensional Lie algebra $\stackrel{\circ }{\bf g%
}$ \cite{h}. It means that $\stackrel{\circ }{\bf g}={\bf k}\stackrel{.}{+}%
{\bf p}$ as a linear space, and

\begin{equation}
\label{4.1}\left[ {\bf k},{\bf k}\right] \subset {\bf k},\left[ {\bf k},{\bf %
p}\right] \subset {\bf p}{\em ,}\left[ {\bf p},{\bf p}\right] \subset {\bf k.%
}
\end{equation}

so that ${\bf k}$ is a subalgebra in $\stackrel{\circ }{\bf g}$. We suppose
that the decomposition $\stackrel{\circ }{\bf g}={\bf k}\stackrel{.}{+}{\bf p%
}$ is orthogonal with respect to the standard scalar product on $\stackrel{%
\circ }{\bf g},$ so that the projectors $P_{{\bf p}}$ and $P_{{\bf k}}$ onto
the subspaces ${\bf p}$ and ${\bf k}$ are orthogonal. Let $\widetilde{%
\stackrel{\circ }{\bf g}}=C^\infty \left( S^1,\stackrel{\circ }{\bf g}%
\right) $ and let us introduce the left currents $l_x,l_t\in $ $\widetilde{%
\stackrel{\circ }{\bf g}}$,

\begin{equation}
\label{4.2}l_x=g^{-1}\partial _xg,l_t=g^{-1}\partial _tg,g\in C^\infty
\left( S^1,\stackrel{\circ }{\em G}\right) =\widetilde{\stackrel{\circ }{\em %
G}},
\end{equation}

where $\stackrel{\circ }{\em G}$ is a Lie group corresponding to $\stackrel{%
\circ }{\bf g}$. Let $A_\mu ,B_\mu $ be the two projections of the current, :

\begin{equation}
\label{4.3}A_\mu =P_{{\bf k}}l_\mu ,B_\mu =P_{{\bf p}}l_\mu .
\end{equation}

In these notations the action of the nonlinear sigma model has the form \cite
{fs}:

\begin{equation}
\label{4.4}
\begin{array}{c}
S\left( g\right) =\frac 12\int tr\left( B_xB_x-B_tB_t\right) dxdt, \\
S\left( g\right) \in Fun\left( \widetilde{\stackrel{\circ }{\em G}}\right) .
\end{array}
\end{equation}

The action functional is unchanged under the right gauge action of the group
$\widetilde{{\em K}}=C^\infty \left( S^1,{\em K}\right) $:

\begin{equation}
\label{4.5}g\mapsto gk,k\in \widetilde{{\em K}},g\in \widetilde{\stackrel{%
\circ }{\em G}},
\end{equation}

so that this is a well defined function on the symmetric space $\widetilde{%
\stackrel{\circ }{\em G}}/$ $\widetilde{{\em K}}$. The equations of motion
which follow from the action (\ref{4.4}) are:

\begin{equation}
\label{4.6}
\begin{array}{c}
\partial _tB_t+\left[ A_t,B_t\right] =\partial _xB_x+\left[ A_x,B_x\right] ,
\\
\partial _xA_t-\partial _tA_x+\left[ A_x,A_t\right] +\left[ B_x,B_t\right]
=0, \\
\partial _xB_t-\partial _tB_x+\left[ A_x,B_t\right] -\left[ A_t,B_x\right]
=0.
\end{array}
\end{equation}

The two last equations are zero curvature conditions which serve to restore
the group variable $g\in $ $\widetilde{\stackrel{\circ }{\em G}}/$ $%
\widetilde{{\em K}}$ from $A_\mu ,B_\mu $.

For this model there exists a Lax pair:

\begin{equation}
\label{4.7}
\begin{array}{c}
L=-\left( A_x+\frac \lambda 2\left( B_x+B_t\right) +\frac 1{2\lambda }\left(
B_x-B_t\right) \right) , \\
T=-\left( A_t+\frac \lambda 2\left( B_x+B_t\right) -\frac 1{2\lambda }\left(
B_x-B_t\right) \right) .
\end{array}
\end{equation}

The equations of motion (\ref{4.6}) are expressed as the zero curvature
condition:

\begin{equation}
\label{4.8}\left[ \partial _x-L,\partial _t-T\right] =0.
\end{equation}

Now we define a natural Poisson structure connected with the Lax pair (\ref
{4.7}). Let ${\bf g}=\stackrel{\circ }{\bf g}\otimes {\bf C}\left[ \lambda
,\lambda ^{-1}\right] $ be the current Lie algebra with the scalar product (%
\ref{3.3}). Let ${\bf g}^\sigma $ be the twisted current Lie algebra which
corresponds to the \ Cartan automorphism $\sigma $ associated with $\left(
{\bf k,p}\right) $:

\begin{equation}
\label{4.9}\sigma \mid _{{\bf k}}=id,\sigma \mid _{{\bf p}}=-id.
\end{equation}

Thus

\begin{equation}
\label{4.10}{\bf g}^\sigma =\left\{ X\left( \lambda \right) \in {\bf g}%
:\sigma X\left( -\lambda \right) =X\left( \lambda \right) \right\} .
\end{equation}

For our model we may choose the standard r-matrix Lie-Poisson structure
bracket (\ref{2.9}) using the r-matrix

\begin{equation}
\label{4.11}r=P_{+}-P_{-};
\end{equation}

where $P_{-}$ is the projection operator onto the negative part of a Laurent
series and $P_{+}$ is the complementary projection operator . The projection
operator $P_{-}$ has the kernel:

\begin{equation}
\label{4.12}P_{-}\left( \lambda ,\mu \right) =t_A\frac{\mu ^2}{\lambda
^2-\mu ^2}+t_B\frac{\lambda \mu }{\lambda ^2-\mu ^2},
\end{equation}

where $t_A=\left( P_{{\bf k}}\otimes P_{{\bf k}}\right) t,t_B=\left( P_{{\bf %
p}}\otimes P_{{\bf p}}\right) t$ are the ${\bf k-}$ and ${\bf p-}$
components of the Casimir element, respectively. The r-matrix (\ref{4.11})
has a symmetric part with the kernel:

\begin{equation}
\label{4.13}s\left( \lambda ,\mu \right) =t_A.
\end{equation}

\begin{example}
{\bf The principal chiral field.}
\end{example}

Let $\stackrel{\circ }{\bf g}={\bf a}\oplus {\bf a}$ be the direct sum of
two copies of a simple Lie algebra ${\bf a}$. Let ${\bf k}$ be the diagonal
subalgebra in $\stackrel{\circ }{\bf g}$ and ${\bf p}$ be the anti-diagonal
subspace:

\begin{equation}
\label{4.14}
\begin{array}{c}
{\bf k}=\left\{ \left( x,x\right) ,x\in {\bf a}\right\} , \\ {\bf p}=\left\{
\left( x,-x\right) ,x\in {\bf a}\right\} , \\ \text{so that corresponding
automorphism }\sigma \text{ is }\sigma \left( x,y\right) =\left( y,x\right).

\end{array}
\end{equation}

This is a Riemannian symmetric pair. The symmetric space for the
corresponding current group is $\widetilde{{\em A}}\times \widetilde{{\em A}}%
/\widetilde{{\em A}}$, where $\widetilde{{\em A}}$ is the current group of
the Lie group corresponding to the Lie algebra ${\bf a}$. In this case $%
\widetilde{{\em A}}\simeq \widetilde{{\em P}}$ and acts on $\widetilde{{\em A%
}}\times \widetilde{{\em A}}$ according to (\ref{4.5}):

$$
\begin{array}{c}
\widetilde{{\em A}}\times \left( \widetilde{{\em A}}\times \widetilde{{\em A}%
}\right) \rightarrow \widetilde{{\em A}}\times \widetilde{{\em A}}, \\
g\circ \left( g_1,g_2\right) \mapsto \left( g_1g,g_2g\right)
\end{array}
$$

So in this case the action (\ref{4.4}) is a well-defined function on $%
\widetilde{{\em A}}$. If we define the projection
\begin{equation}
\label{4.15}
\begin{array}{c}
\pi :
\widetilde{{\em A}}\times \widetilde{{\em A}}\rightarrow \widetilde{{\em A}}%
, \\ \pi :\left( g_1,g_2\right) \mapsto g_1g_2^{-1}=g
\end{array}
\end{equation}

onto the quotient space $\widetilde{{\em A}}\simeq \widetilde{{\em A}}\times
\widetilde{{\em A}}/\widetilde{{\em A}}$ and define the currents

\begin{equation}
\label{4.16}l_\mu ^k=g_k^{-1}\partial _\mu g_k,k=1,2,l_\mu =g^{-1}\partial
_\mu g,
\end{equation}

then the variables $A_\mu $ and $B_\mu $ have the form:

\begin{equation}
\label{4.17}
\begin{array}{c}
A_\mu =\left( \frac 12\left( l_\mu ^1+l_\mu ^2\right) ,\frac 12\left( l_\mu
^1+l_\mu ^2\right) \right) , \\
B_\mu =\left( \frac 12\left( l_\mu ^1-l_\mu ^2\right) ,-\frac 12\left( l_\mu
^1-l_\mu ^2\right) \right) ,
\end{array}
\end{equation}

and the action (\ref{4.4}) takes the form:

\begin{equation}
\label{4.18}S\left( g\right) =\int \left( l_xl_x-l_tl_t\right) dxdt.
\end{equation}

This is the action of the principal chiral field model on $\widetilde{{\em A}%
}$, so that we deal with a realization of this model on the symmetric space $%
\widetilde{{\em A}}\times \widetilde{{\em A}}/\widetilde{{\em A}}$ .The Lax
operator (\ref{4.7}) is a direct sum of the two ones:

\begin{equation}
\label{4.19}L\left( \lambda \right) =\left( L^1\left( \lambda \right)
,L^1\left( -\lambda \right) \right) ,
\end{equation}

and the r-matrix (\ref{4.11}) is a direct sum of a two copies of the
rational r-matrix:

\begin{equation}
\label{4.20}r=\left(
\begin{array}{cc}
P_{+}-P_{-} & 0 \\
0 & P_{+}-P_{-}
\end{array}
\right) .
\end{equation}

Let us return to the situation of example 3.1. Our model is a direct sum of
two copies of this example. The quantum lattice algebra is a direct sum of
two copies of the algebra (\ref{3.11}), where $L^n$ is a quantum version of
a lattice approximation of the Lax operator $L^1\left( \lambda \right) $. If

\begin{equation}
\label{4.21}M\left( \lambda \right) =\left( M^1\left( \lambda \right)
,M^1\left( -\lambda \right) \right)
\end{equation}

is the quantum monodromy matrix for our model, then $M^1\left( \lambda
\right) $ satisfies the relation (\ref{3.12}). If ${\bf a}=sl\left( 2\right)
$ then the generating function for the integrals of motion is:

\begin{equation}
\label{4.22}tr_qM^1\left( \lambda \right) +tr_qM^1\left( -\lambda \right) ,
\end{equation}

and we may apply the technique of the modified Bethe Ansatz developed in
section 3 to calculate the spectrum.

\begin{example}
{\bf \ The n-field.}
\end{example}

Let $\stackrel{\circ }{\bf g}=su(2)$ and let ${\bf k}=\stackrel{\circ }{\bf h%
}$ be the Cartan subalgebra of $su(2)$, and ${\bf p}$ the complementary
subspace. This is a Riemannian symmetric pair and the corresponding
automorphism $\sigma $ is given by (cf. \ref{3.32}):

\begin{equation}
\label{4.23}\sigma X=DXD^{-1},X\in su(2),D=\left(
\begin{array}{cc}
1 & 0 \\
0 & -1
\end{array}
\right) .
\end{equation}

Let

\begin{equation}
\label{4.24}
\begin{array}{c}
A_x=\left(
\begin{array}{cc}
ia & 0 \\
0 & -ia
\end{array}
\right) ,B_x=\left(
\begin{array}{cc}
0 & A_1+A_{-1} \\
-\left( \overline{A}_1+\overline{A}_{-1}\right) & 0
\end{array}
\right) , \\
B_t=\left(
\begin{array}{cc}
0 & A_1-A_{-1} \\
-\left( \overline{A}_1-\overline{A}_{-1}\right) & 0
\end{array}
\right) ,
\end{array}
\end{equation}

then in this case the Lax operator (\ref{4.7}) has the form:

\begin{equation}
\label{4.25}L=-\left(
\begin{array}{cc}
ia & \lambda A_1+\frac 1\lambda A_{-1} \\
-\lambda \overline{A}_1-\frac 1\lambda \overline{A}_{-1} & -ia
\end{array}
\right) .
\end{equation}

Since the automorphism $\sigma $ coincides with the Coxeter automorphism (%
\ref{3.32}) for $sl(2)$, the twisted affine Lie algebra ${\bf g}^\sigma $ (%
\ref{4.10}) coincides with the compact real form of the twisted realization
of the algebra $A_1^{\left( 1\right) }$. To define a Poisson structure we
may use the r-matrix (\ref{4.11}) which in this case is precisely the
r-matrix from example 3.1 So that we are returning to the situation
investigated in this example. The lattice quantum algebra is given by
theorem 3.4, part 2, where $L^n$ is a quantum version of a lattice
approximation of the Lax operator (\ref{4.25}). According to theorem 3.3 for
this algebra there exists an ultralocalization. Let

\begin{equation}
\label{4.26}
\begin{array}{c}
\widehat{L}^n\left( \lambda \right) =-\left(
\begin{array}{cc}
iA_0^n & \lambda A_1^n+\frac 1\lambda A_{-1}^n \\
-\lambda A_1^{n^{*}}-\frac 1\lambda A_{-1}^{n^{*}} & iA_0^{n^{*}}
\end{array}
\right) , \\
G^n=\left(
\begin{array}{cc}
a_n & 0 \\
0 & a_n^{*}
\end{array}
\right) ,
\end{array}
\end{equation}

where we use the notations of theorem 3.3 and $*$ is the anti-involution in
this algebra (it is supposed that $\mid q\mid =1$).We shall give an explicit
realization of this algebra via the canonical Weyl pairs

\begin{equation}
\label{4.27}
\begin{array}{c}
U_i^nV_i^n=qV_i^nU_i^n,i=1,2;n=1,\ldots ,N, \\
U_i^{n^{*}}=U_i^{n^{-1}},V_i^{n^{*}}=V_i^{n^{-1}}.
\end{array}
\end{equation}

We omit the complicated calculations and give only the result:

\begin{equation}
\label{4.28}
\begin{array}{c}
A_0^n=\left( 1-q^{-1}V_1^{n^{-2}}\right) U_1^nU_2^n, \\
A_1^n=V_1^nU_2^n,A_{-1}^n=V_1^{n^{-1}}U_2^n, \\
a_n=V_1^{n^{-\frac 12}}V_2^{n^{\frac 14}}V_2^{n+1^{-\frac 14}}.
\end{array}
\end{equation}

The Lax operator $\widehat{L}^n\left( \lambda \right) $ coincides up to an
unessential factor with the one of the Lattice Sine-Gordon model \cite{f3}:

\begin{equation}
\label{4.29}\widehat{L}^n=L_{LSG}^n\otimes U_2^n.
\end{equation}

Since $U_2^n$ is the shift operator in the standard Weyl representation of
the algebra (\ref{4.27}), we may choose the reference state in the form:

\begin{equation}
\label{4.30}
\begin{array}{c}
\Omega =\Omega _{LSG}\otimes
{\bf 1}, \\ {\bf 1}=\stackunder{N}{\underbrace{1\otimes \ldots \otimes 1}},
\end{array}
\end{equation}

where $\Omega _{LSG}$ is the reference state of the Lattice Sine-Gordon
model. U$_2^n$ acts trivially on $1$, so that the spectrum of our model
coincides with the one for the Lattice Sine-Gordon model. Here we are using
the gauge equivalence of these two models and the connection between
generating functions for the integrals of motion:

\begin{equation}
\label{4.31}
\begin{array}{c}
trM=q^{\frac 12}tr
\widehat{M}, \\ M=L^1\cdot \ldots \cdot L^N, \\
\widehat{M}=\widehat{L}^1\cdot \ldots \cdot \widehat{L}^N.
\end{array}
\end{equation}

\section{Conclusion}

In this section we formulate some problems related to the subject of this
paper.

The first problem is connected with a generalization of our construction of
nonultralocal algebras to an arbitrary graph. It is not difficult to
construct a nonultralocal algebra on a graph using a solution of the
Yang-Baxter equation on the square of some Lie algebra and the 'polyuble
construction' \cite{fr1}. In the finite-dimensional case these algebras were
introduced in \cite{fr1} in connection with the quantization of the moduli
space. The lattice algebra on a graph similar to one considered in example 1
of this paper was investigated in \cite{a}.

A more interesting problem is the regularization of the Poisson bracket
relations for the monodromy $\ $ for a more wide class of scalar products on
current algebras, e.g. when

$$
\left( X\left( \lambda \right) ,Y\left( \lambda \right) \right) =Res\text{ }%
tr\left( X\left( \lambda \right) Y\left( \lambda \right) \right) \phi \left(
\lambda \right) d\lambda ,
$$

where $\phi \left( \lambda \right) $ is some rational function. The
Belavin-Drinfeld classification theorem does not allow to define
regularization in the way considered in section 2, because there exist no
nontrivial solutions of the Classical Yang-Baxter equation on ${\bf g}\oplus
{\bf g}$ of the type (\ref{2.19}) which are skew-symmetric with respect to
this scalar product . One of the possible ways to tackle with this situation
is to use Quasi-Hopf algebras \cite{d3}. This requires the study of
r-matrices which do not satisfy the Yang-Baxter equation. However, the
physical motivation to introduce such algebras is not clear.

Another problem is the construction of representations of nonultralocal
algebras. These representations are important for the definition of
reference states $\Omega $ and of the Bethe Ansatz construction. A natural
way to solve this problem is the ultralocalization procedure. As we have
seen, ultralocalization does not exist for all lattice algebras. It is
interesting to describe all representations of nonultralocal algebras and to
choose from them those which arise from an ultralocalization procedure. For
finite-dimensional algebras this question is connected with the anomaly
problem in the quantum field theory \cite{afs}.

The last problem is to construct so called fundamental Lax operator for
nonultralocal algebras \cite{f3}. The first step to its solution was made in
\cite{fv1}.

The authors would like to thank  L.D.Faddeev, F.A.Smirnov, and A.Yu.Alekseev
 for numerous helpful discussions.

\end{document}